\documentclass[iop,apj,twocolumn,numberedappendix,appendixfloats]{openjournal}
\usepackage{graphicx} 
\usepackage{amsmath}	
\usepackage{bm}
\usepackage{hyperref}
\usepackage{pythonhighlight}
\usepackage{fontawesome5}
\usepackage{hyperref}
\usepackage{tikz}
\usepackage{float}
\usepackage{orcidlink}

\hypersetup{breaklinks,colorlinks,citecolor=blue,linkcolor=blue,urlcolor=blue}
\makeatletter
\newcommand{\github}[1]{%
   \href{#1}{\faGithubSquare}%
}
\makeatother
\newcommand{\meer}{\textsc{meer21cm}}
\newcommand{\hi}{H\textsc{i}}
\newcommand{\txteq}[1]{\,{#1}\,}

\newcommand{\secref}[1]{\hyperref[#1]{Section~\ref*{#1}}}
\newcommand{\appref}[1]{\hyperref[#1]{Appendix~\ref*{#1}}}
\begin{document}

\title{\meer: an analysis pipeline and comprehensive toolkit for \hi\ intensity mapping}

\author{
    Zhaoting Chen \orcidlink{0000-0002-4965-8239}$^1$\footnote{zhaoting.chen@roe.ac.uk}, 
    Steven Cunnington \orcidlink{0000-0001-6594-107X}$^{2}$, 
    Daniel Tassie \orcidlink{0009-0000-1440-2578}$^{2}$,
    Alkistis Pourtsidou$^{1,3}$,
    Laura Wolz$^{4}$,
    Gabriele Autieri$^{5}$,
    Matilde Barberi-Squarotti\orcidlink{0009-0007-8964-5807}$^{6,7,8,9}$,
    Jos\'e Luis Bernal\orcidlink{0000-0002-0961-4653}$^{10}$,
    Phil Bull$^{4,11}$,
    Stefano Camera\orcidlink{0000-0003-3399-3574}$^{9,12,13,11}$,
    Isabella P.\ Carucci \orcidlink{0000-0001-5287-0065}$^{14,15}$,
    Brandon Engelbrecht$^{11}$,
    Jos\'e Fonseca\orcidlink{0000-0003-0549-1614}$^{16,17,11}$,
    Karin Fornazier\orcidlink{0000-0003-0578-9533}$^{11}$,
    Keith Grainge\orcidlink{0000-0002-6780-1406}$^{4}$,
    Jiakang Han$^{9}$,
    Wenkai Hu \orcidlink{0000-0002-3108-5591}$^{18,11}$,
    Melis O. Irfan\orcidlink{0000-0003-2021-7357}$^{19}$,
    Piyanat Kittiwisit$^{11}$,
    Yichao Li$^{20}$,
    Sefa Pamuk\orcidlink{0009-0004-0852-8624}$^{10}$,
    Mario G. Santos$^{11,21}$,
    Marta Spinelli\orcidlink{0000-0003-0148-3254}$^{22,11}$,
    Jingying Wang\orcidlink{0000-0002-5598-2668}$^{23,11}$,
    Amadeus Witzemann$^{11,4}$, 
    Boyan Zhao$^{18}$
    \\(MeerKLASS Collaboration)
}
\affiliation{%
$^1$Institute for Astronomy, The University of Edinburgh, Royal Observatory, Edinburgh EH9 3HJ, UK\\
$^2$Institute of Cosmology \& Gravitation, University of Portsmouth, Dennis Sciama Building,
Portsmouth, PO1 3FX, UK\\
$^3$Higgs Centre for Theoretical Physics, School of Physics and Astronomy, Edinburgh EH9 3FD, UK\\
$^4$Jodrell Bank Centre for Astrophysics, Department of Physics \& Astronomy, The University of Manchester, Manchester M13 9PL, UK
$^5$SISSA -- International School for Advanced Studies, Via Bonomea 265, 34136 Trieste, Italy\\
$^{6}$Dipartimento di Fisica, Universit\`a degli Studi di Milano, via G.\ Celoria 16, 20133 Milano, Italy\\
$^{7}$INFN -- Istituto Nazionale di Fisica Nucleare, Sezione di Milano, via G.\ Celoria 16, 20133 Milano, Italy\\
$^{8}$INAF -- Istituto Nazionale di Astrofisica, Osservatorio Astrofisico di Brera-Merate, via Brera 28, 20121 Milano, Italy\\
$^{9}$Dipartimento di Fisica, Universit\`a degli Studi di Torino, via P.\ Giuria 1, 10125 Torino, Italy\\
$^{10}$Instituto de F{\'i}sica de Cantabria (IFCA), CSIC-Univ. de Cantabria, Avda. de los Castros s/n, E-39005 Santander, Spain\\
$^{11}$Department of Physics \& Astronomy, University of the Western
Cape, Robert Sobukwe Road, Cape Town 7535, South Africa\\
$^{12}$INFN -- Istituto Nazionale di Fisica Nucleare, Sezione di Torino, via P.\ Giuria 1, 10125 Torino, Italy\\
$^{13}$INAF -- Istituto Nazionale di Astrofisica, Osservatorio Astrofisico di Torino, 10025 Pino
Torinese, Italy\\
$^{14}$INAF -- Istituto Nazionale di Astrofisica, Osservatorio Astronomico di Trieste, Via G.B.\ Tiepolo 11, 34131 Trieste, Italy\\
$^{15}$IFPU -- Institute for Fundamental Physics of the Universe, Via Beirut 2, 34151 Trieste, Italy\\
$^{16}$Instituto de Astrof\'isica e Ci\^encias do Espa\c{c}o, Universidade do Porto CAUP, 4150-762 Porto, Portugal\\
$^{17}$Departamento de F\'isica e Astronomia, Faculdade de Ci\^{e}ncias, Universidade do Porto, Rua do Campo Alegre 687, 4169-007, Portugal\\
$^{18}$State Key Laboratory of Radio Astronomy and Technology, National Astronomical Observatories, CAS, A20 Datun Road, Beijing 100101, China\\
$^{19}$Institute of Astrophysics, University of Cambridge, Madingley Road, CB3 0HA\\
$^{20}$Department of Physics, College of Sciences, Northeastern University, Wenhua Road, Shenyang 11089, China\\
$^{21}$South African Radio Astronomy Observatory (SARAO), Liesbeek House, Cape Town, 7700, South Africa\\
$^{22}$Observatoire de la C$\hat{\mathrm{o}}$te d’Azur, Laboratoire Lagrange, Bd de l’Observatoire, CS 34229, 06304 Nice cedex 4, France\\
$^{23}$Shanghai Astronomical Observatory, Chinese Academy of Sciences, 80 Nandan Road, Shanghai 200030, China 
}

\thanks{$^\star$E-mail: \href{mailto:zhaoting.chen@roe.ac.uk}{zhaoting.chen@roe.ac.uk}}
\thanks{$^\star$E-mail: \href{mailto:steve.cunnington@port.ac.uk}{steve.cunnington@port.ac.uk}}
\journalinfo{The Open Journal of Astrophysics}
\begin{abstract}
We present \meer, a comprehensive \textsc{python} package for cosmological data analysis of
single-dish \hi\ intensity mapping surveys.
This package is simple to use, with a modularised code structure designed for interactive usage.
\meer\ is designed for data analysis, with particular focus on the UHF-band observation of MeerKAT Large Area Synoptic Survey (MeerKLASS).
We explicitly impose \meer\ to be survey-oriented, ensuring consistent modelling of
observational effects in the clustering power spectrum with the survey specifications and
data analysis choices.
\meer\ covers a large range of data analysis procedures post calibration, 
including data read-in, foreground cleaning, power spectrum estimation,
mock simulation, transfer function corrections and parameter inference.
It handles both \hi\ intensity maps and overlapping galaxy catalogues,
allowing for multi-tracer and cross-correlation analysis between MeerKLASS and optical galaxy surveys.
Tested with a simulated survey of ten $750\,$deg$^2$ sky patches in the redshift sub-band $0.6\,{<}\,z\,{<}\,0.8$,
the \meer\ pipeline achieves per-cent accuracy in the power spectrum estimation for $k \in [0.02, 0.2]\,{h{\rm Mpc}^{-1}}$,
with deviations $\lesssim 0.5\sigma$ between the mock and the model power spectra, where $\sigma$ is the signal variance.
The \meer\ package is publicly available and easy to install, with a comprehensive documentation website
at \url{https://meer21cm.readthedocs.io}.
\end{abstract}

\section{Introduction}
\label{sec:intro}
Neutral hydrogen (\hi) intensity mapping \citep{1997ApJ...475..429M,2001JApA...22...21B,2004MNRAS.355.1339B,2008PhRvL.100i1303C,2009MNRAS.397.1926W,2010Natur.466..463C} has become an emerging probe of the cosmic large-scale structure over the past decade.
By mapping the distribution of the flux density of the 21\,cm emission line of neutral hydrogen, intensity maps observed through radio telescopes can be used to probe the distribution of galaxies in the post-reionization Universe, and therefore the properties of the underlying dark matter.
In particular, at relatively low redshifts ($z\lesssim 3$), a cosmological 21\,cm survey can be performed by using the auto-correlation of the voltage signal, commonly referred to as the ``single dish'' mode \citep{2013MNRAS.434.1239B}, to achieve large survey volumes with low angular resolution maps.
For the next generation of radio telescopes, such as the Square Kilometre Array Observatory (SKAO), intensity mapping surveys will be able to measure precisely the clustering signal of \hi, providing constraints on the cosmological model that are competitive to well-established probes such as galaxy clustering \citep[][and Advancing Astrophysics II SKAO science book\footnote{\href{https://www.skao.int/en/science-users/557/advancing-astrophysics-ii}{skao.int/en/science-users/557/advancing-astrophysics-ii}}]{2020PASA...37....7S}.

Experiments targeting the cosmic \hi\ intensity mapping signal have made substantial progress.
Measurements of the \hi\ clustering power spectrum have been made in cross-correlation with spectroscopic optical galaxies, using the Green Bank Telescope \citep{2013ApJ...763L..20M,2013MNRAS.434L..46S,2022MNRAS.510.3495W}, the Parkes telescope \citep{2018MNRAS.476.3382A}, and the MeerKAT telescope \citep{2023MNRAS.518.6262C,2025A&A...703A.222C,2025MNRAS.537.3632M}. 
The \hi\ signal has also been detected by stacking intensity maps onto overlapping galaxy coordinates with MeerKAT \citep{2025ApJS..279...19C} and the Canadian Hydrogen Intensity Mapping Experiment \citep{2023ApJ...947...16A}. Lastly, exciting progress has been made in claimed detections of the \hi\ power spectrum in intensity mapping auto-correlation, again using MeerKAT \citep{2023arXiv230111943P, 2026arXiv260223055T,MatildeAutoPk} and CHIME \citep{2025arXiv251119620C}.

In this paper, the context in which we discuss our work is centred around the MeerKAT Large Area Synoptic Survey (MeerKLASS; \citealt{2016mks..confE..32S}) using the MeerKAT telescope.
After conducting two ${\sim}\,200$\,deg$^2$ single-patch surveys in MeerKAT's L-band \citep{2021MNRAS.505.3698W,2025MNRAS.537.3632M}, which provided early validation of the single-dish technique (see \citealt{2026Ap&SS.371...16C,Cunnington:2026opd} for recent reviews of these results), MeerKLASS is now conducting an observational campaign in UHF-band ($580\,{<}\,\nu\,{<}\,1000$\,MHz), which will reach a wider redshift range ($0.4\,{<}\,z\,{<}\,1.45$) and span ${\sim}\,10{,}000$\,deg$^2$.
Since MeerKAT is the precursor to, and will be part of, the mid-frequency array of the SKAO, our work will also directly serve future SKAO intensity mapping surveys.

21\,cm experiments face unique challenges, dealing with various systematic effects.
The emission line signal is intrinsically weak and several orders of magnitude fainter than the radio foregrounds.
Component separation techniques are needed to isolate the cosmological signals (e.g. \citealt{2014MNRAS.441.3271W, 2016ApJS..222....3Z, 2016MNRAS.456.2749O,2019AJ....157....4Z,2020MNRAS.499..304C,2021MNRAS.508.3551I,2025A&A...703A.222C,Spinelli:2026mfo}), which rely on the fact that the foregrounds are spectrally smooth and the cosmic \hi\ is not.
As a result, effects that perturb the spectral structure of the signal will induce systematics into the data.
In particular, calibration errors due to incomplete sky models \citep{2016MNRAS.461.3135B, 2020MNRAS.494.5018H}, polarisation leakage \citep{2014MNRAS.444.3183A, 2020MNRAS.499..304C, 2021MNRAS.504..208C} and beam chromaticity \citep{2021MNRAS.506.5075M,2024arXiv241209527S,2025ApJS..279...19C} can be sources of these systematics.

To mitigate the systematics and to accurately reconstruct the underlying cosmological signal, \hi\ intensity mapping requires power spectrum analysis that differs from conventional clustering analysis.
For example, the foreground removal procedure removes part of the \hi\ signal, and requires additional corrections to the estimated power spectrum \citep{2015ApJ...815...51S}.
These corrections are typically computed numerically by constructing mock observations \citep{2023MNRAS.523.2453C}, a process that has been shown to be equivalent to reconstructing the power spectrum window functions \citep{2025MNRAS.542L...1C}.
The mock simulation must follow the survey specifications, since the \hi\ clustering amplitude depends on the survey due to its anisotropy.
The measured \hi\ signal is highly anisotropic for a number of reasons.
First, single dish 21\,cm experiments typically have very high resolution along the line-of-sight and low angular resolution (see e.g. \citealt{2015ApJ...803...21B}), leading to anisotropic smoothing effects.
Second, the signal loss due to foreground removal is predominantly along the large line-of-sight scales, in low radial wavenumber $k_\parallel$.
Third, observational effects such as the beam smoothing and the redshift space distortions produce highly anisotropic clustering signal along the transverse $\bm{k}_\perp$ and radial $k_\parallel$ directions (see e.g. \citealt{2020MNRAS.496..415C}).
Finally, the intensity map cube needs to be sampled onto a 3-dimensional Cartesian grid, with anisotropic compensations to correct for re-gridding effects and pixelization that depend on the survey geometry \citep{2024MNRAS.528.5586C}.
As a result, intensity mapping data analysis is often highly complex.
Choosing a specific data cut or a particular foreground removal strategy means that all steps of the data analysis are affected, from power spectrum estimation, window function reconstruction, to model inference. 
Meanwhile, as an emerging probe, robust understandings of \hi\ intensity mapping analysis are nascent, and each dataset requires different pipeline settings, such as the number of foreground modes removed. It often requires visual checks and manual testing on the data, creating a significant challenge to scale up the data analysis procedure to larger survey volumes.

\hi\ intensity mapping is an inherently spectroscopic probe, with excellent redshift overlap with optical galaxies. To utilise the scientific potential, cross-correlation with spectroscopic galaxy catalogues is necessary for the data analysis. The cross-correlation poses further challenges of consistently performing \hi\ auto-power, galaxy auto-power and \hi-galaxy cross-power estimation, all within one pipeline. Correspondingly, simulations with consistent \hi\ and galaxy mocks need to be included in the pipeline, requiring additional validation.

As \hi\ intensity mapping enters the SKAO era in the near future, it is clear that robust routines are needed to address the challenges in data analysis and increased data volume. 
In this work, we present \meer, developed to be the core analysis pipeline for imminent MeerKLASS observations, while also establishing a scalable foundation for future SKAO campaigns\footnote{In principle, by changing the rest-frame frequency of the emission line, any 21\,cm pipeline could be trivially repurposed for other line intensity mapping surveys \citep[see e.g.][]{2022A&ARv..30....5B,2022ApJ...927..161K,2023A&A...676A..62V,2026ApJ...999..177L}.}.
It is built on the foundation of previous analysis codes used for the MeerKLASS L-band data analysis \citep[see e.g.][]{2023MNRAS.518.6262C}. 
To meet the demands of future \hi\ intensity mapping analysis, our pipeline has been overhauled and extended to include the following features:
\begin{itemize}
    \item \textbf{Modularised}. \meer\ uses the \textsc{python} class infrastructure and its inheritance feature to modularise the different steps of data analysis. As we outline in \secref{sec:code}, we define modules of the package in a way that naturally arises from the data analysis procedure, from survey data file loading to parameter inference. Each module is therefore easy to use and modify, making \meer\ highly adaptable and extendable.
    \item \textbf{Survey-oriented}. Contrary to theoretical calculations that typically start with cubic boxes, with Cartesian comoving dimensions, and modelling the signal in $(|\bm{k}|,\mu \txteq{=}k_\parallel /|\bm{k}| )$ space, all calculations in \meer\ start by defining a survey with the sky area and frequency range. 
    The power spectrum is always modelled at the 3D $\bm{k}$-vector level, and the 1D average is performed by including the exact 3D $\bm{k}$-modes as the survey volume rectangular grid.  This ensures realistic modelling of \hi\ signal and observational effects.
    \item \textbf{Consistent}. As mentioned, for \hi\ intensity mapping, any change to data cuts, foreground removal, weighting or any other technical detail will have non-negligible impact on all aspects of the data analysis, from mock simulations of transfer function calculations to theoretical modelling of the power spectrum. We implement a caching system in \meer\ that explicitly tracks the dependencies of various quantities on different data analysis settings. Combined with the survey-oriented settings, \meer\ ensures all calculations consistently account for the observational effects and automatically handles any changes to the analysis, as we demonstrate in \secref{sec:code}.
    \item \textbf{Easy-to-use}. We provide a documentation website\footnote{\url{https://meer21cm.readthedocs.io}} with comprehensive API summary and a large number of examples. A full data analysis pipeline for the MeerKLASS L-band is also presented. As it is highly modularised and intuitive to interact with, users can easily plug in their own data analysis steps and test out different details, such as studying the optimal strategy of foreground removal, while keeping the rest of the pipeline intact and consistent.
    \item \textbf{Comprehensive}. \meer\ is designed to cover all data analysis steps post-calibration, and provides a comprehensive toolkit to generate mock observations, theoretical modelling and parameter inference, on top of the power spectrum estimation. The functionality and framework we present in this paper will continuously be used and improved in future development.
    \item \textbf{Rigorous}. \meer\ is tested and validated, with unit-test coverage of $\sim 100 \%$ of the code. Apart from the unit tests, end-to-end validations across the different aspects of the pipeline are also performed, as we describe in \secref{sec:validation}.
    \item \textbf{Open}. \meer\ is publicly available and easy to install. We envision that the release of \meer\ with the accompanying documentation will draw interest among the intensity mapping community. The extensive use cases provided by the community will lead to further improvement and continuous development.
\end{itemize}

The rest of this paper is structured as follows.
In \secref{sec:formalism}, we review the power spectrum estimator used in \hi\ intensity mapping data analysis, as well as the corresponding theoretical modelling.
In \secref{sec:mock}, we describe the simulation routine for generating mocks that are used for validating the power spectrum estimation pipeline.
An end-to-end data analysis pipeline and its validations are presented in \secref{sec:validation}.
The power spectrum and mock simulation formalism will then motivate the code structure described in \secref{sec:code}.
Also in \secref{sec:code}, we provide a few use cases that demonstrate the power and flexibility of \meer.
We discuss the outlook for \meer\ towards cosmology with the SKAO in \secref{sec:future}, and conclude in \secref{sec:conclusion}.
Throughout this paper, we adopt the $\Lambda$CDM model of cosmology reported in \cite{2020A&A...641A...6P}.

\section{Power Spectrum Formalism}
\label{sec:formalism}

For an \hi\ density field $\rho_{\rm \hi}(\bm{x})$ at redshift $z$, the brightness temperature distribution of the field, $T_{\rm \hi}(\bm{x})$, follows \citep{2006PhR...433..181F} 
\begin{equation}
    T_{\rm \hi}(\bm{x}) = C_{\rm \hi} \rho_{\rm \hi}(\bm{x})=\frac{3A_{12}h_{\rm P}c^3(1+z)^2}{32\pi m_{\rm H}k_{\rm B} \nu_{21}^2 H(z)} \rho_{\rm \hi}(\bm{x}),
\end{equation}
where $h_{\rm P}$ is the Planck constant, $k_{\rm B}$ is the Boltzmann constant, $m_{\rm H}$ is the mass of the hydrogen atom, $A_{12}$ is the emission coefficient of the 21-cm line transmission, $\nu_{21}\,{\sim}\,1420\,$MHz is the rest frequency of the 21-cm emission \citep{4313902} and $H(z)$ is the Hubble parameter at redshift $z$.

The Fourier transform of the brightness temperature field is defined as
\begin{equation}
    \tilde{T}_{\rm \hi}(\bm{k}) = \int \frac{{\rm d^3}\bm{x}}{\rm V} T_{\rm \hi}(\bm{x})\, {\rm exp}\big[-i\bm{k}\bm{x} \big].
\end{equation}

In the linear Kaiser regime \citep{1987MNRAS.227....1K} supplemented with a phenomenological Finger-of-God (FoG; \citealt{1972MNRAS.156P...1J}) correction, the \hi\ brightness temperature power spectrum in redshift space, $P_{\rm \hi}(\bm{k})$, can be written as
\begin{equation}
    P_{\rm \hi}(\bm{k}) = {\rm V}\langle|\tilde{T}_{\rm \hi}(\bm{k})|^2  \rangle = \bar{T}_{\rm \hi}^2 \frac{ (b_{\rm \hi} + f \mu^2)^2}{1 + |\bm{k}|^2\mu^2(\sigma_{\rm p}^{\rm \hi})^2} P_{\rm m}(\bm{k}),
\label{eq:himodel}
\end{equation}
where $\bm{k}$ is the 3D wavenumber vector, $\rm V$ is the survey volume, $\bar{T}_{\rm \hi}$ is the average brightness temperature of 21\,cm line emission, $b_{\rm \hi}$ is the linear \hi\ bias, $f$ is the growth rate of the Universe, $\mu = k_\parallel / |\bm{k}|$ with $k_\parallel$ being the wavenumber along the line-of-sight, $\sigma_{\rm p}^{\rm \hi}$ is the velocity dispersion parameter in the unit of comoving length, and $P_{\rm m}(\bm{k})$ is the linear matter power spectrum.
All quantities in \autoref{eq:himodel}, except $\mu$ and $\bm{k}$, are redshift dependent, and we omit the notation of $z$-dependency for simplicity.

The observed brightness temperature field, on the other hand, has several additional observational effects.
In general, the observed \hi\ brightness temperature field can be written as
\begin{equation}
\begin{split}
    T_{\rm obs}(\bm{x}) = \Bigg[&\bigg[\Big[\big[{T}_{\rm \hi} (\bm{\theta},\nu)  \circledast B(\bm{\theta},\nu)\big] 
 \times {w}^f_{\rm \hi}(\bm{\theta},\nu)\Big]
 \circledast G_{\rm map} \bigg] \\ &
 \otimes R(\nu)
 \circledast G_{\rm grid}\Bigg](\bm{x}) 
\end{split}
\label{eq:tobs}
\end{equation}
where $\bm{\theta}$ denotes the position vector on the sky, 
$\nu$ denotes the observing frequency,  
$\circledast$ denotes convolution, 
$B$ denotes the telescope beam, 
${w}^f_{\rm \hi}(\bm{\theta},\nu)$ denotes the \textit{field}-level weights which we will discuss later, 
$\otimes$ denotes tensor product, 
$G_{\rm map}$ is the convolution kernel induced by the map-making process, 
$R$ is the foreground removal operator, 
$G_{\rm grid}$ is the convolution kernel induced by the gridding of map data to rectangular grids.

The power spectrum estimator for the \hi\ signal is then
\begin{equation}
    \hat{P}_{\rm \hi}(\bm{k}) = \frac{\rm V}{\big\langle (w^{f}_{\rm \hi}w^g_{\rm \hi})^2\big\rangle_{\rm V}}\,\Big|\mathcal{F}\big[w^g_{\rm \hi}{T}_{\rm obs}\big](\bm{k})\Big|^2,
\label{eq:phiest}
\end{equation}
where $w^g_{\rm \hi}$ is the \textit{grid}-level weighting for the \hi\ temperature field, $\mathcal{F}$ denotes Fourier transformation, $\langle \rangle_{\rm V}$ denotes the volume average, and we apply a rescaling of $\big\langle (w^{f}_{\rm \hi}w^g_{\rm \hi})^2\big\rangle_{\rm V}$ to normalise the amplitude of the estimator \citep{2010MNRAS.406..803B,2019MNRAS.489..153B}.

To perform the gridding procedure, a fiducial cosmology is assumed to convert the sky coordinates $(\bm{\theta},\nu)$ to comoving coordinates $\bm{x}$.
A mismatch between the true cosmology and the fiducial cosmology creates the distortion of the dimensions of the clustering, known as the Alcock-Paczy\'nski (AP; \citealt{1979Natur.281..358A}) effect.
The observed $\bm{k}$-mode, based on fiducial cosmology, is distorted compared to the true $\bm{k}'$-mode, leading to a mismatch of the measured clustering scales as well as the clustering amplitude. While the AP effect is fully implemented in \meer, it is not relevant to the validation tests which we discuss later. 
Therefore, we omit the AP effect terms in the power spectrum formalism. Detailed study of cosmological inference using \meer\ will be discussed in follow-up work, and we will later demonstrate the inference routine by fitting bias and \hi\ density parameters in \autoref{apdx:parameterinference}.

The expectation value of the power spectrum estimator can then be \emph{approximated} as
\begin{equation}
\begin{split}
    \langle & \hat{P}_{\rm \hi}(\bm{k}) \rangle  \approx  P^{\rm model}_{\rm \hi}(\bm{k}) \\
    = & \bigg[\Big[P_{\rm \hi} |\tilde{B}|^2 \big|\tilde{G}_{\rm map}\big|^2 \big|\tilde{G}_{\rm grid}\big|^2 \Big] \circledast \Big[\Big| \mathcal{F}\big[{w^f_{\rm \hi}{w^g_{\rm \hi}}\big]}\Big|^2  \Big]\bigg](\bm{k})\\
    &\times \mathcal{T}(\bm{k}) \bigg/ \Big\langle (w^{f}_{\rm \hi}w^g_{\rm \hi})^2\Big\rangle_{\rm V},
\end{split}
\label{eq:phiobsmodel}
\end{equation}
where $P_{\rm \hi}$ is the theoretical \hi\ power spectrum described in \autoref{eq:himodel}, $\tilde{B}$, $\tilde{G}_{\rm map}$ and $\tilde{G}_{\rm grid}$ denote the Fourier transform of the kernels described in \autoref{eq:tobs}, and $\mathcal{T}$ is the foreground transfer function that corrects for the signal loss due to foreground removal, i.e. reversing the distortion caused by $R(\nu)$. 
Comparing with \autoref{eq:tobs}, we can see that \autoref{eq:phiobsmodel} assumes that their smoothing effect is localised, i.e. the widths of the map-making and gridding kernels in $\bm{k}$-space are negligible compared to the field-level weights, so that the multiplication of $w^f_{\rm \hi}$ is commutable with the convolution of $G_{\rm map}$ and $G_{\rm grid}$. Furthermore, it is assumed that the effect of foreground removal is also commutable with the rest of the operations, and can be described by a transfer function.

To perform cross-correlations with a galaxy catalogue survey, we must also specify the formalism for this analysis. For the galaxy number overdensity field $\delta_{\rm gal}(\bm{x})$, the clustering power spectrum is
\begin{equation}\label{eq:galmodel}
    P_{\rm gal}(\bm{k}) = {\rm V}\langle|\tilde{\delta}_{\rm gal}(\bm{k})|^2  \rangle =  \frac{ (b_{\rm gal} + f \mu^2)^2}{1 + |\bm{k}|^2\mu^2(\sigma_{\rm p}^{\rm gal})^2} P_{\rm m}(\bm{k}),
\end{equation}
Similar to \autoref{eq:tobs}, the observed galaxy number density field can be written as
\begin{equation}
    {\rm n}_{\rm obs}(\bm{x}) = \Bigg[\Big[{\rm n}_{\rm gal}
 \times {w}^f_{\rm n}\Big]
 \circledast G_{\rm grid}\Bigg](\bm{x}),
\label{eq:gobs}
\end{equation}
where ${\rm n}_{\rm gal}$ is the underlying galaxy number density, ${w}^f_{\rm n}$ is the field-level weights for the galaxy field. Since the galaxy number density field is directly obtained from a discrete catalogue, there is no beam smoothing or map-making effect in ${\rm n}_{\rm obs}(\bm{x})$. The observed galaxy number overdensity field is then mean-centered so that
\begin{equation}
    \delta^{\rm n}_{\rm obs}(\bm{x}) = \frac{{w}^g_{\rm n}(\bm{x})\,{\rm n}_{\rm obs}(\bm{x})}{\big\langle {w}^g_{\rm n}(\bm{x})\, {\rm n}_{\rm obs} (\bm{x}) \big\rangle_{\rm V}} - 1,
\end{equation}
where $w^g_{\rm n}(\bm{x})$ is the grid-level weights for the galaxy field.
The power spectrum estimator for the galaxy clustering is then
\begin{equation}
    \hat{P}_{\rm gal}(\bm{k}) = \frac{{\rm V} \big(\big\langle w^{f}_{\rm n}w^g_{\rm n}\big\rangle_{\rm V}\big)^2 }{\big\langle (w^{f}_{\rm n}w^g_{\rm n})^2\big\rangle_{\rm V}}\,\big|\tilde{\delta}^{\rm n}_{\rm obs}(\bm{k})\big|^2.
\label{eq:pgest}
\end{equation}
Compared to \autoref{eq:phiest}, the galaxy auto-power estimator has an extra factor of $\big(\big\langle w^{f}_{\rm n}w^g_{\rm n}\big\rangle_{\rm V}\big)^2$ arising from the mean-centering. A brief derivation is presented in \appref{apdx:galaxyps} for reference.

The observed galaxy power spectrum can then be approximately modelled as
\begin{equation}
\begin{split}
    \langle & \hat{P}_{\rm gal}(\bm{k}) \rangle  \approx  P^{\rm model}_{\rm gal}(\bm{k}) \\
    = & \bigg[\Big[P_{\rm gal}\big|\tilde{G}_{\rm grid}\big|^2 \Big] \circledast \Big[\Big| \mathcal{F}\big[{w^f_{\rm n}{w^g_{\rm n}}\big]}\Big|^2  \Big]\bigg](\bm{k})\bigg/ \Big\langle (w^{f}_{\rm n}w^g_{\rm n})^2\Big\rangle_{\rm V} \\
    & + P_{\rm SN}(\bm{k}),
\end{split}
\end{equation}
where $P_{\rm SN}(\bm{k})$ is the shot noise, which we assume to be Poissonian, and follows
\begin{equation}
    P_{\rm SN}(\bm{k}) = \frac{C_1(\bm{k})}{\langle {\rm n}_{\rm obs} \rangle_{\rm V}} \frac{\big\langle (w^g_{\rm n})^2 \big\rangle_{\rm N}}{\big\langle \big(w^g_{\rm n}w^f_{\rm n}\big)^2 \big\rangle_{\rm V}\big( \big\langle w^g_{\rm n} \big\rangle_{\rm N} \big)^2},
\label{eq:psn}
\end{equation}
where $C_1(\bm{k})$ denotes the gridding window for the shot noise \citep{2005ApJ...620..559J}, and $\langle \rangle_{\rm N}$ denotes the galaxy number weighted average, so that for any field $f$, 
\begin{equation}
    \langle f\rangle_{\rm N} = \langle f\, {\rm n}_{\rm obs}\rangle_{\rm V} / \langle  {\rm n}_{\rm obs}\rangle_{\rm V}.
\end{equation}
A derivation of the shot noise is presented in \appref{apdx:galaxyps}.

Finally, the theoretical \hi-galaxy cross-power is modelled as
\begin{equation}
    P_\times (\bm{k}) = \bar{T}_{\rm \hi} \frac{ r_\times b_{\rm \hi}b_{\rm gal}+ f \mu^2b_{\rm \hi} + f \mu^2b_{\rm gal}  + f^2 \mu^4}{\sqrt{\big[1 + |\bm{k}|^2\mu^2(\sigma_{\rm p}^{\rm \hi})^2\big] \big[1 + |\bm{k}|^2\mu^2(\sigma_{\rm p}^{\rm gal})^2\big]}} P_{\rm m}(\bm{k}),
\end{equation}
where $r_\times$ is the cross-correlation coefficient between the two tracers.

The estimator for the cross-power is
\begin{equation}
    \hat{P}_{\times} = \frac{{\rm V} \big\langle w^{f}_{\rm n}w^g_{\rm n}\big\rangle_{\rm V} }{\big\langle w^{f}_{\rm n}w^g_{\rm n}w^{f}_{\rm \hi}w^g_{\rm \hi}\big\rangle_{\rm V}}\,\Big|\big[\tilde{\delta}^{\rm n}_{\rm obs}(\bm{k})\big]
    \big[\tilde{T}_{\rm obs}(\bm{k})\big]^*\Big|_{\rm Re}.
\label{eq:pcest_cross}
\end{equation}
The expectation value of the estimator can be modelled as
\begin{equation}
\begin{split}
    \langle & \hat{P}_{\times}(\bm{k}) \rangle \approx P_{\times}^{\rm model}(\bm{k})  \\
    = & \bigg[\Big[P_{\times} |\tilde{B}| \big|\tilde{G}_{\rm map}\big| \big|\tilde{G}_{\rm grid}\big|^2 \Big] \circledast 
    \Big[\mathcal{F}\big[{w^f_{\rm \hi}{w^g_{\rm \hi}}\big]} \Big]\Big[\mathcal{F}\big[{w^f_{\rm n}{w^g_{\rm n}}\big]} \Big]^*
    \bigg](\bm{k})\\
    &\times \mathcal{T}(\bm{k}) \bigg/ \Big\langle w^{f}_{\rm \hi}w^g_{\rm \hi}w^{f}_{\rm n}w^g_{\rm n}\Big\rangle_{\rm V}.
\end{split}
\label{eq:pcrossmodel}
\end{equation}
Note that, for both the \hi-auto power and the cross-power, the foreground removal effect is corrected by the same factor $\mathcal{T}(\bm{k})$.
This is specific to foreground cleaning using Principal Component Analysis (PCA), as shown in \cite{2023MNRAS.523.2453C} and \cite{2025MNRAS.542L...1C}.

For the mock simulation and validation tests shown later, we need to choose the fiducial parameters for the input model as well as the functional form for the kernels such as $\tilde{B}$ and $\tilde{G}_{\rm map}$. The model choices are summarised in \autoref{tab:model}.

\section{Mock Simulation}
\label{sec:mock}

In this section, we describe the mock simulation routine implemented in \meer.
The survey characteristics we describe below are fixed for consistency with the validation tests presented later in \secref{sec:validation}. These choices aim to approximately emulate a single unmasked \textit{patch} from the ongoing MeerKLASS UHF observation campaign, in which multiple patches have now been amassed (also discussed later in \secref{sec:validation}).

We simulate the mock \hi\ survey with a sky area of $\rm {\sim}\,750\,deg^2$ at $125\,{<}\,{\rm R.A.}\,{<}\,175\,$deg and ${-}10\,{<}\,{\rm Dec.}\,{<}\,5\,$deg. 
We assign a frequency range of $789.11\,{<}\,\nu\,{<}\,887.53\,$MHz, corresponding to an approximate redshift range of $0.6\,{<}\,z\,{<}\,0.8$, 
with an effective redshift of $z_{\rm eff}\,{=}\,0.7$. 
We choose this redshift range as it corresponds to the ``LRG2'' samples of the Dark Energy Spectroscopic Instrument (DESI) survey \citep{2025JCAP...04..012A}, 
and cross-correlation with DESI is one of the immediate science goals for MeerKLASS. 
We set the frequency channel resolution to $\delta\nu\txteq{\sim}0.133$\,MHz, consistent with the MeerKAT UHF receivers. 
The sky map is defined in the \textsc{astropy} World Coordinate System (WCS) format\footnote{\href{https://docs.astropy.org/en/stable/wcs/}{docs.astropy.org/en/stable/wcs/}}.
\meer\ also supports the definition of the sky map in the \textsc{healpix} format\footnote{\href{https://healpy.readthedocs.io/en/stable/}{healpy.readthedocs.io}}, 
and the choice of the format gives negligible differences for the results shown in this paper.
The angular size of the sky map pixel is set to be $\rm (0.5\, deg)^2$, consistent with the map-making choice currently adopted for MeerKLASS UHF observations.

With the survey specifications, we generate the mock \hi\ signal with the following procedure. Using the survey area and the redshift range, a minimum enclosing rectangular box in Cartesian comoving space is found for the survey lightcone.
In our case, the Cartesian box size is [716.8, 2348.2, 824.5]\,Mpc. 
We assign this to a \textit{simulation} grid\footnote{There are two sets of Cartesian grids in our pipeline, a high-resolution grid for simulating the density fields and a lower-resolution grid which the sky map is sampled onto for performing power spectrum estimation. The former is referred to as the ``simulation grid'' and the latter the ``estimation grid''.} with higher resolution, $65\times 215\times 2073$ cells of size [11.3, 11.2, 0.4]\,Mpc, corresponding to approximately half of the pixel size and frequency channel bandwidth. 
Using the input power spectrum of \autoref{eq:himodel}, a log-normal \hi\ density field is generated across the Cartesian simulation grid following \cite{2011MNRAS.416.3017B}. This \hi\ density field is then multiplied by the average brightness temperature, $\bar{T}_{\rm \hi}$. This differs from the approach of other simulation packages \citep[e.g.][]{2023ApJ...958....4L}, which Poisson sample galaxies from the matter distribution, assigns positions and velocities to them, and then luminosities according to some input luminosity function. The motivation for our approach is computational speed, but our framework is designed so that any simulation can be injected into the pipeline and follow the subsequent analysis features.
We discuss the details of our lognormal implementation in \appref{apdx:lognormal}.

To create the \textit{sky map} of \hi\ temperature fluctuations, the cell values from the Cartesian simulation are gridded into ($\boldsymbol{\theta},\nu$) voxels on the estimation grid. 
For each map voxel, the temperature value is taken to be the mean value of the Cartesian cells that fall into that voxel.
As with the input Cartesian simulated grid, any alternative sky-based map (e.g. \textsc{healpix}) can be injected into the pipeline at this stage, allowing $N$-body lightcone mocks to become the cosmological input \citep[see e.g.][]{Ronconi:2026jij}.
At this point, the sky map can have additional observational effects added (detailed in the following sub-section). These are optional and in this work we clearly indicate what we are including depending on the demonstration. Finally, based on the survey area and frequency range, map voxels outside the survey lightcone are masked.

\meer\ also has the capability to simulate a coherent overlapping galaxy catalogue for investigating cross-correlations. Similarly to the \hi\ intensity maps, these begin with a log-normal \textit{galaxy} density field being generated from the input galaxy power spectrum (\autoref{eq:galmodel}).
The density field is then Poisson sampled to generate the discrete galaxy positions following an input redshift distribution.
The Cartesian coordinates for the discrete galaxy positions are then converted to ($\boldsymbol{\theta},z$) sky coordinates.
Any galaxies falling outside the survey lightcone or selection function are excluded from the final catalogue.

\subsection{Observational contaminants}

\begin{figure*}
    \centering
    \includegraphics[width=1\linewidth]{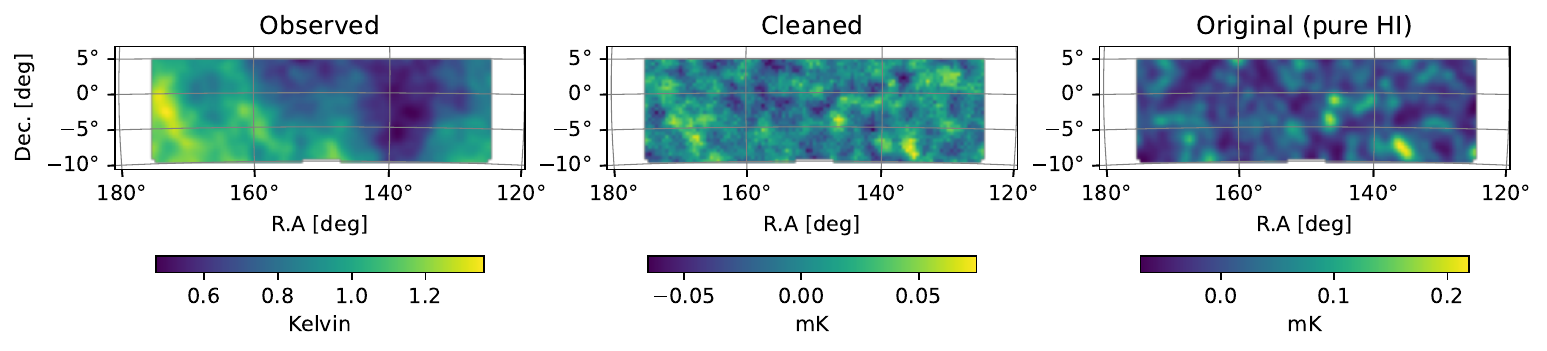}
    \caption{Mock map demonstrations of the effects from observational contamination. 
    Left panel shows the ``observed'' map before cleaning illustrating the dominant foregrounds. 
    The telescope beam and thermal noise have also been included in this observed map. 
    This is then cleaned by removing 5 PCA modes to give the central panel which drops the amplitude of the fluctuations by several orders of magnitude to reveal thermal noise and cosmological \hi. 
    The structure of the latter is revealed by comparison with the pure \hi\ map shown by the right panel. 
    The maps in each panel have been averaged along the 10 central frequency channels. \github{https://github.com/meerklass/meer21cm/blob/main/papers/validation/map_illus.ipynb}}
    \label{fig:maps_demo}
\end{figure*}

To emulate observational contamination in the data, the mocks can also include foreground contamination, convolution with telescope's beam pattern, and additive thermal noise. To simulate the foreground contamination we add in maps produced from the Python Sky Model \citep{2017MNRAS.469.2821T,2021JOSS....6.3783Z, 2025ApJ...991...23P}, with default model settings [\texttt{s1,f1,d1,a1,c1}]\footnote{See \href{https://pysm3.readthedocs.io/en/latest/}{pysm3.readthedocs.io} for more details.} to include synchrotron, free-free, dust, AME, and CMB sky contributions. The \textsc{healpy} outputs are interpolated onto the same \textsc{astropy} WCS maps as the mock \hi\ data. An example of this is shown in the left panel of \autoref{fig:maps_demo} (this includes the telescope beam and thermal noise, discussed next).

The sky map is then convolved with the beam, which in our mock tests we assume to be Gaussian; however, \meer\ has functionality for more complex and realistic beam patterns, e.g. \textsc{katbeam}\footnote{\href{https://github.com/ska-sa/katbeam}{github.com/ska-sa/katbeam}}. 
The Gaussian beam's full-width-half-maximum (FWHM) is computed with 
\begin{equation}
    \theta_\text{FWHM} = \frac{c}{\nu \, D_\text{dish}},
\end{equation}
where $D_\text{dish}\txteq{=}13.5$\,m is the diameter of the MeerKAT dish.
The maps are convolved with this Gaussian kernel, and all panels in \autoref{fig:maps_demo} include this to emulate the beam.

After the beam convolution, we add a Gaussian thermal noise realisation with zero mean and RMS of $\sigma_{\rm N}\txteq{=}0.04\,$mK, arbitrarily chosen for these demos to be similar to the \hi\ fluctuations at zero lag (in reality the noise will be ${\sim}\,10$ times higher). This can be more robustly set to emulate a survey's particular subset of data by following the radiometer equation,
\begin{equation}
    \sigma_{\rm N}(\boldsymbol{\theta},\nu) = \frac{T_{\rm sys}(\boldsymbol{\theta},\nu)}{\sqrt{2\delta\nu \, \delta t \, N_{\rm hits}(\boldsymbol{\theta},\nu)}},
\label{eq:sigmaN}
\end{equation}
where $T_{\rm sys}$ is the system temperature for the observations, $\delta t$ is the integrated time for each telescope reading \textit{hit} (typically 2\,sec for MeerKLASS), and $N_{\rm hits}$ is the amassed number of hits within each map voxel.
This noise is added onto the observed map (\autoref{fig:maps_demo} left panel) but is sub-dominant relative to the foregrounds for our chosen setup (also true for integrated MeerKLASS observations), hence is only evident in the foreground cleaned map in the middle panel. 
We note that, since we focus primarily on validating the accuracy of the power spectrum estimator, we do not include the thermal noise simulation for the rest of the paper for simplicity.
In actual data analysis, cross-correlating different datasets of the same survey area is typically used to eliminate the noise bias in the \hi\ auto-power (e.g. \citealt{2025MNRAS.541..476M}). 
We discuss the chosen foreground cleaning process in the following section.

\subsection{Blind foreground cleaning}

We adopt a Principal Component Analysis (PCA)-based blind foreground cleaning method to remove the foreground contamination. Given our observed intensity map, $m(\boldsymbol{\theta},\nu)$ which has been mean-centred in each frequency channel, with weights $w(\boldsymbol{\theta},\nu)$, the frequency-frequency covariance is calculated as
\begin{equation}\label{eq:freq_cov}
    \textbf{\textsf{C}}_{\nu_1 \nu_2} = \frac{\sum\limits_{i=1}^{N_\theta}\left[(w m)\left(\boldsymbol{\theta}_i, \nu_1\right) \times(w m)\left(\boldsymbol{\theta}_i, \nu_2\right)\right] }{ \sum\limits_{i=1}^{N_\theta}\left[w\left(\boldsymbol{\theta}_i, \nu_1\right) w\left(\boldsymbol{\theta}_i, \nu_2\right)\right]}\,,
\end{equation}
where $i$ iterates over each pixel in the angular plane and $N_\theta$ is the total
number of pixels. The eigenmodes [$\boldsymbol{v}_1$,$\boldsymbol{v}_2$,...,$\boldsymbol{v}_{N_\nu}$], ranked from the largest eigenvalue to the smallest, are obtained from the eigendecomposition of $\textbf{\textsf{C}}_{\nu_1 \nu_2}$. A choice is made for the number of eigenmodes, $N_{\rm fg}$, that are deemed to contain the dominant and frequency-correlated foreground contamination. From this, we form the PCA matrix 
\begin{equation}\label{eq:PCAmatrix}
    \textbf{\textsf{R}}^{\mathrm{PCA}} = \textbf{\textsf{I}}-\sum_{n=1}^{N_{\mathrm{fg}}} \boldsymbol{v}_n \boldsymbol{v}_n^{\mathrm{T}},
\end{equation}
where $\textbf{\textsf{I}}$ is the identity matrix.

\begin{equation}
    m_{\text {clean }}\left(\boldsymbol{\theta}, \nu_i\right)=\sum_j \textbf{\textsf{R}}_{i j}^{\mathrm{PCA}} \, m\left(\boldsymbol{\theta}, \nu_j\right) \,,
\label{eq:pca}
\end{equation}
where $j$ loops over each frequency channel and $_{ij}$ denotes the $i^{\rm th}$ row and $j^{\rm th}$ column of the PCA matrix. 

The resulting cleaned map is shown by the central panel of \autoref{fig:maps_demo} where $N_{\rm fg}\txteq{=}5$ modes have been removed. 
The vast drop in amplitude relative to the original observed map (left panel), and comparison with the pure-\hi\ map (right panel) reveals the recovered structure, with the differences coming from the thermal noise and signal loss from the PCA cleaning, plus some small residual foregrounds.
While the choice of $N_{\rm fg}=5$ is optimistic in actual data analysis,
we note that improvements in data quality have driven the number of PCA modes to be lower,
as seen in e.g. \cite{2025A&A...703A.222C}.
We use $N_{\rm fg}=5$ for the mock simulation to showcase the pipeline, 
and leave a more detailed discussion for future work.

\subsection{Regridding for power spectrum analysis}

\begin{figure}
    \centering
    \includegraphics[width=\linewidth]{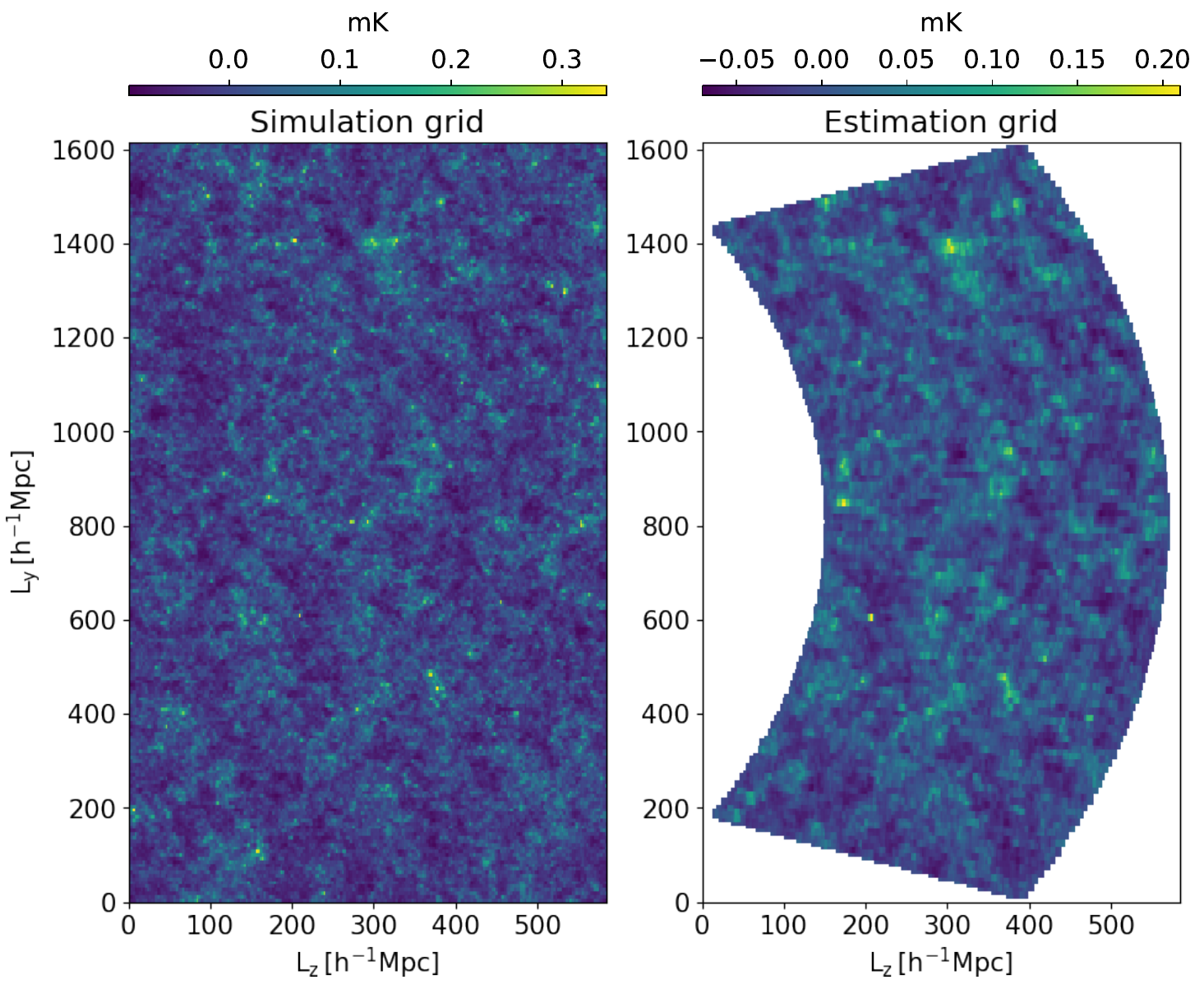}
    \caption{Comparison between input and output gridded fields in comoving Cartesian space. 
    The left panel shows the generated input mock on the simulation grid. After the mapping step, this is then interpolated onto the estimation grid (right panel) for power spectrum analysis. White space is not covered by the survey footprint which follows a lightone, traced by the geometry of the angular and redshift coverage.  
    This example shows only cosmological \hi, i.e. no foregrounds, beam or noise. 
    Both panels have been averaged along the $L_{\rm x}$ direction. \github{https://github.com/meerklass/meer21cm/blob/main/papers/validation/map_illus.ipynb}}
    \label{fig:logrealisation_map_grid}
\end{figure}

The resulting \hi\ intensity map and overlapping galaxies, both in sky coordinates, represent the observable data products from the respective surveys. As in a real power spectrum data analysis pipeline, these are then required to be transformed back to the Cartesian comoving space \citep{2024MNRAS.528.5586C}. They are assigned onto a lower resolution \textit{estimation grid}, with $21{\times}71{\times}695$ cells and uniform cell sizes of [34.1, 33.1, 1.19]\,Mpc.
We choose the cloud-in-cell (CIC) mass assignment scheme to interpolate the observed \hi\ intensity map and galaxy positions.
An illustration of this routine for a \hi-only field (without any observational effects) is shown in \autoref{fig:logrealisation_map_grid} (right panel). The transformational changes to resolution and survey mask are demonstrated by comparison with the input \textit{simulation grid} (left panel).

\subsection{Transfer function correction}
\label{subsec:TFcorrection}
The map data is modified by the PCA foreground cleaning procedure, as described in \autoref{eq:pca}.
The effects of PCA cleaning are then corrected by a transfer function.
To calculate the transfer function, we use mock signal injection following \cite{2023MNRAS.523.2453C}.
A mock \hi\ signal $m_{\rm mock}$ is generated and injected into the original map data, giving $m + m_{\rm mock}$.
The PCA matrix is then calculated following \autoref{eq:freq_cov} and \autoref{eq:PCAmatrix} but with $m + m_{\rm mock}$ as the input. The cleaned mock residual is then defined as 
\begin{equation}
    m_{\text {clean }}\left(\boldsymbol{\theta}, \nu_i\right)=\sum_j \textbf{\textsf{R}}_{i j}^{\mathrm{PCA}} \, m_{\rm mock}\left(\boldsymbol{\theta}, \nu_j\right) \,,
\end{equation}
i.e. the PCA modes are projected from just the mock. Previous versions have projected out modes from $m + m_{\rm mock}$. This will provide identical results, albeit with more variance. 
The residual cleaned mock is gridded onto the Cartesian estimation grid, as is the original uncleaned mock \hi\ signal (or a mock galaxy catalogue if a cross-correlation). The transfer function is defined as the cross-correlation between the cleaned and uncleaned mock signal,
\begin{equation}
    \mathcal{T}(k) = \frac{\langle \mathcal{P}[m_{\rm mock},m_{\rm clean}]\rangle_{\rm m}}{\langle \mathcal{P}[m_{\rm mock},m_{\rm mock}]\rangle_{\rm m}},
\label{eq:TF}
\end{equation}
where $\langle \rangle_{\rm m}$ denotes averaging across mock realisations, and $\mathcal{P}[,]$ denotes the operator that calculates the cross-power between two tracer fields. 

We note that it is different from the transfer function defined in \cite{2023MNRAS.523.2453C}, namely in the order of taking the ratio and the ensemble average.
The choice is to match the numerical transfer function calculation with the analytical derivation of \cite{2025MNRAS.542L...1C}.

\subsection{Model Inference}
\label{subsec:modelinference}
The measured power spectra can then be used to perform Bayesian inference of the underlying parameters.
For this work, we adopt mock-based covariance, where the covariance of the data vector is 
\begin{equation}
    \mathbf{C} = \langle \bm{d}_m \bm{d}^{\rm T}_m \rangle_{\rm m}
\label{eq:cov}
\end{equation}
where $\bm{d}_m$ is the mock data vector and $\langle \rangle_{\rm m}$ denotes the average over the realisations.
Given the measured power spectrum data vector and the estimated covariance $\mathbf{C}$, the log-likelihood for a given parameter set $\bm{\vartheta}$ is
\begin{equation}
\begin{split}
    \log \mathcal{L}(\bm{\mathcal{M}}|\bm{\vartheta}) = & -\frac{1}{2} \frac{n_{m}-n_d-2}{n_m-1} \frac{1 + B(n_d - n_p)}{1 + A + B(n_p+1)}\\ & \times (\bm{d}-\mathcal{M})^{\rm T}\mathbf{C}^{-1}(\bm{d}-\mathcal{M}),
\end{split}
\label{eq:loglike}
\end{equation}
where $n_m$ is the number of mock realisations when mock-based covariance is used (e.g. \citealt{2025JCAP...04..055F}), $\mathcal{M}$ is the model vector given the parameter set $\bm{\vartheta}$, $n_d$ is the length of the data vector, $n_p$ is the number of parameters, and $A,B$ are correction factors so that
\begin{equation}
    A = \frac{2}{(n_m-n_d-1)(n_m-n_d-4)},
\end{equation}
\begin{equation}
    B = \frac{n_m-n_d-2}{(n_m-n_d-1)(n_m-n_d-4)}.
\end{equation}
The two correction factors to the Gaussian likelihood in \autoref{eq:loglike} correspond to the Hartlap factor \citep{2007A&A...464..399H} and the Percival factor \citep{2014MNRAS.439.2531P}, respectively.

\begin{figure*}[!htbp]
    \centering
    \includegraphics[width=\linewidth]{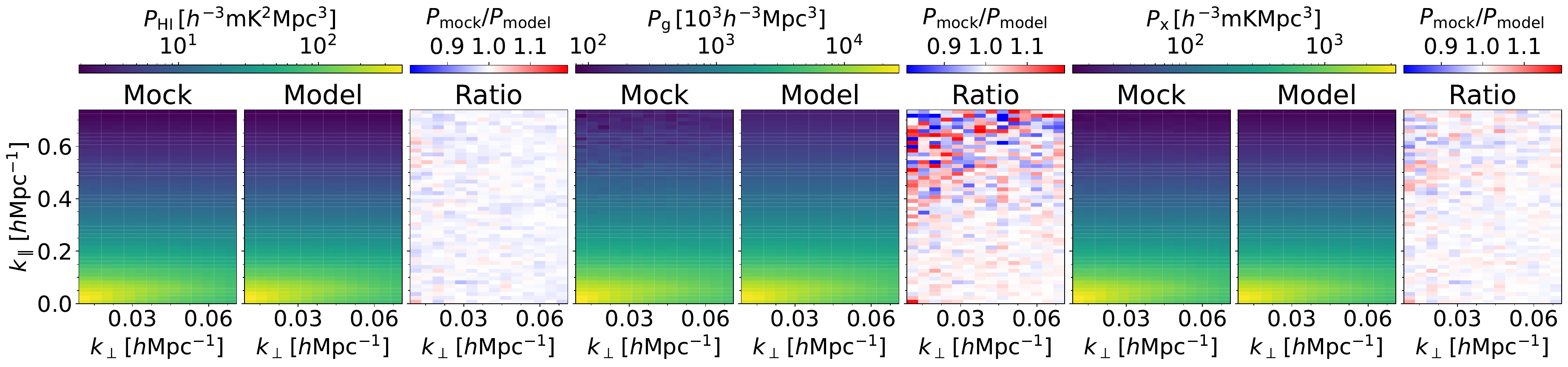}
    \includegraphics[width=\linewidth]{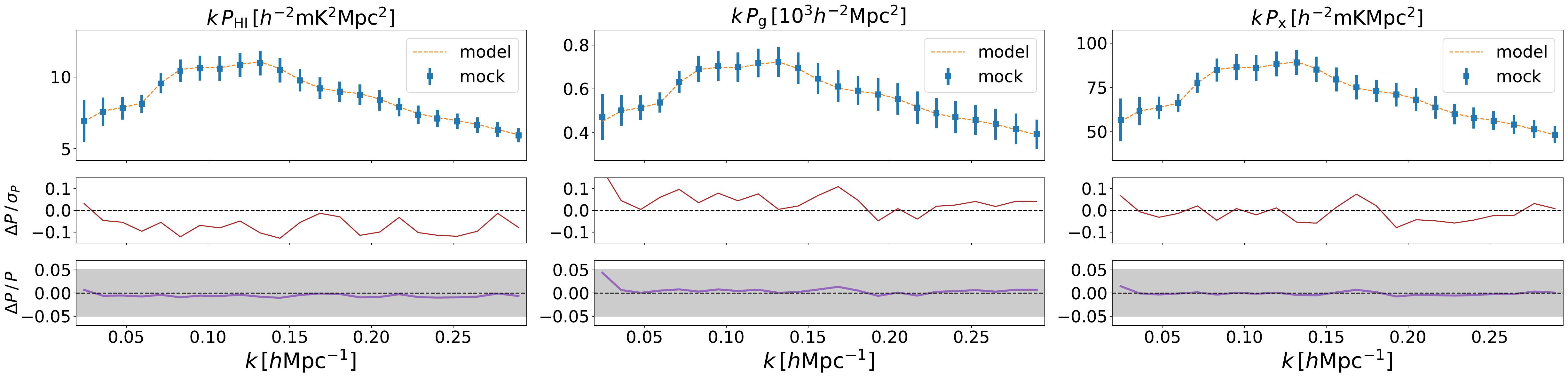}
    
    \caption{
        Validation on the accuracy of the lognormal simulation for the \hi\ signal and the galaxy catalogue (without any observational effects applied). 
        Top panels show the cylindrical power spectra of the \hi\ auto-power, galaxy auto-power, and cross-power, respectively. 
        For each type of the power spectrum, the left panel shows the power spectrum of the mock, the centre panel shows the input model power spectrum, and the right panel shows the ratio between the mock and the model. 
        Results are averaged over 512 realisations. 
        Bottom panels show the results for the 1D power spectra. 
        The top row shows the 1D power spectra. 
        The blue square represents the mock power spectrum, with the error bars representing the standard deviations between the mock realisations. 
        The orange dashed line shows the input model power spectrum. 
        The central row shows the fractional differences between the mock average and the model over the standard deviation. 
        The bottom row shows the fractional difference over the input model, 
        with the shaded region representing the $\pm5\%$ region. \github{https://github.com/meerklass/meer21cm/blob/main/papers/validation/plot_00.ipynb}  }
    \label{fig:mockvalidation}
\end{figure*}

\section{Validation}
\label{sec:validation}
In this section, we validate each step of the \meer\ pipeline by running the fiducial mock simulation described in \secref{sec:mock}.
We generate 512 realisations of the mock to calculate the average mock data vector, as well as the covariance.
Since each realisation simulates the observation for one patch of MeerKLASS UHF data,
we scale the covariance by a factor of $1/N_{\rm patch}$ for results shown in \secref{subsec:mockobservation} and \secref{subsec:transferfunction}, providing a more stringent test of precision.  
$N_{\rm patch}=10$ corresponds roughly to the number of patches with DESI overlap 
in the future preliminary MeerKLASS UHF data analysis.
For simplicity, we showcase the results for the cylindrical power spectra and the 1D power spectrum monopole (a brief study of higher-order multipoles is also presented in \appref{apdx:multipoles}).

The 1D $k$-bins are chosen to be linearly spaced between 0.016 and 0.3 $h{\rm Mpc}^{-1}$ with 24 bins.
For reasons explained later in \secref{subsec:mockobservation},
we employ a $k_x<0.06\,{h{\rm Mpc}^{-1}},\,k_y<0.06\,{h{\rm Mpc}^{-1}}$ cut when averaging the 3D power spectrum to the 1D power spectrum monopole.
A further cut of $k_x>0.015\,{h{\rm Mpc}^{-1}},\,k_y>0.015\,{h{\rm Mpc}^{-1}}$ 
is applied to the 1D power spectrum monopole to exclude the wide-angle modes 
heavily affected by foreground cleaning effects in \secref{subsec:transferfunction}.
We reiterate that the power spectrum is modelled at the full three-dimensional $\bm{k}$-vector level. The averaged one-dimensional power spectrum is obtained only after evaluating the model on the exact discrete set of 3D Fourier modes, defined by the estimator grid. This ensures consistent treatment of model and data.

\subsection{Lognormal simulation}

We first verify the accuracy of the lognormal field simulation, by directly calculating the power spectra on the \hi\ temperature field and the galaxy positions in the simulation grid, before applying any observational effects.
The results are shown in \autoref{fig:mockvalidation}.
In all, we see excellent agreement between the mock and the model, with the differences being less than 0.1$\sigma$ (where $\sigma$ is the standard deviation between the realisations).
The fractional differences of the mock power spectra over the model power spectra are $\sim \pm 1\%$ on all scales of interest.
The results shown in \autoref{fig:mockvalidation} suggest that the lognormal simulation is accurate
and suitable for validating the mock simulation and modelling pipeline.

\subsection{Mock observation}
\label{subsec:mockobservation}
We then propagate the simulated \hi\ fields and galaxy positions to mock sky maps and galaxy catalogues, and apply the beam smoothing, gridding and weighting following the data analysis procedure.
The gridded tracer fields are then used to estimate the power spectra.
In parallel, the model power spectra are modified according to the equations described in \secref{sec:formalism}, to compare against the mock.
This allows us to validate the power spectrum estimation and modelling, excluding the transfer function corrections from foreground removal, which we will include and discuss later.

\begin{figure*}[!htbp]
    \centering
    \includegraphics[width=\linewidth]{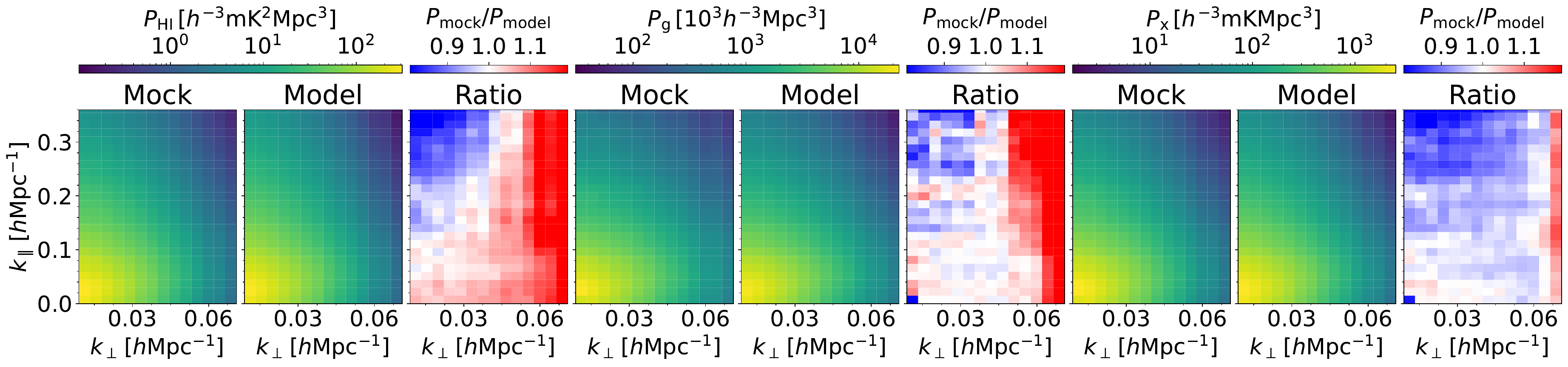}
    \includegraphics[width=\linewidth]{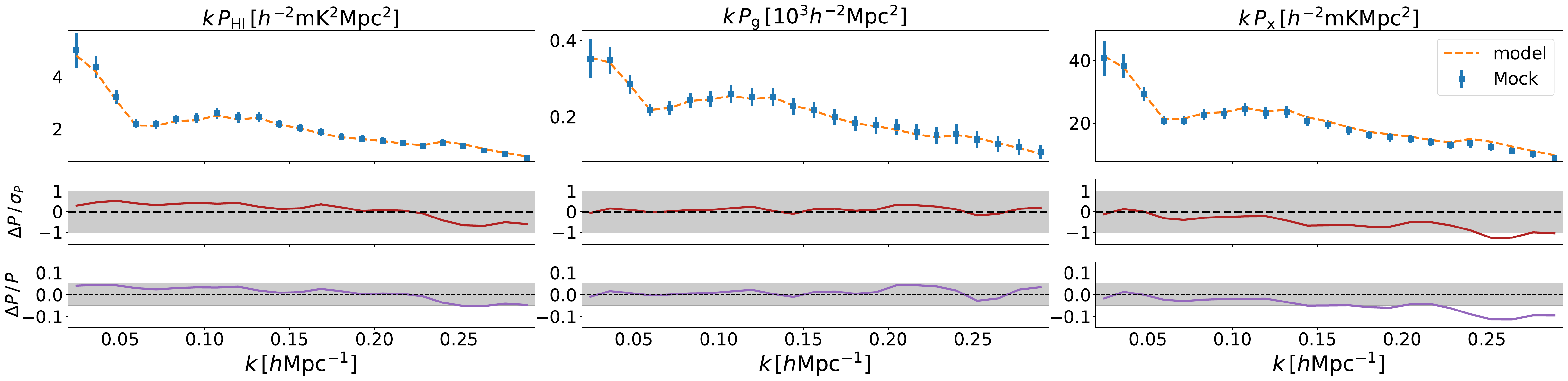}
    
    \caption{
        Validation on the accuracy of the power spectrum estimation from an end-to-end pipeline, without foreground cleaning effects. 
        The top panels show the same types of cylindrical power spectra as \autoref{fig:mockvalidation}, 
        with the mock signal propagated through the pipeline with observational effects. 
        The bottom panels show the 1D power spectra, with the blue square representing the mock power spectrum, 
        and the orange dashed line representing the input model power spectrum. 
        The centre of the bottom panels shows the fractional differences between the mock average and the model over the standard deviation, with the shaded area denoting the $\pm1\sigma$ region. 
        The bottom row shows the fractional difference between the mock average and the model over the model, with the shaded area denoting the $\pm5\%$ region. \github{https://github.com/meerklass/meer21cm/blob/main/papers/validation/plot_01.ipynb} 
    }
    \label{fig:estvalidation}
\end{figure*}

The results on cylindrical power spectra are shown in the top panels of \autoref{fig:estvalidation}.
The top left panels show the average of the mock \hi\ auto-power against the input model.
The cylindrical power spectrum is in agreement with the observational effects applied model
with $\sim 10\%$ accuracy at 
[$|\bm{k}_\perp| \lesssim 0.06\,{h{\rm Mpc}^{-1}}$, $k_\parallel \lesssim 0.2\,{h{\rm Mpc}^{-1}}$].
The decreased accuracy at larger $|\bm{k}_\perp|$ is due to the inaccuracy in modelling the beam size.
Note that, in the mock simulation, the beam size is modelled at each frequency channel,
and directly convolved on the sky map.
On the other hand, in the modelling, we assume a single effective beam size at the effective redshift,
which induces potential bias.
The decreased accuracy at small $|\bm{k}_\perp|$, large $k_\parallel$,
is due to the beyond flat-sky, plane-parallel effects coupled with beam chromaticity.
In our simplified modelling of \autoref{eq:phiobsmodel},
the beam term is an attenuation along the transverse direction.
However, the convolution is performed in the sky map taking into account
the sky projection (emulating real data), whereas the power spectrum estimation is performed assuming plane-parallel.
The mismatch leads to a systematic underestimation of the power spectrum at wide angles and small line-of-sight scales.
We verify the cause of this effect by running the pipeline without beam smoothing effects,
and find that the structure at small $|\bm{k}_\perp|$, large $k_\parallel$
diminishes, as shown in \autoref{apdx:beamsmooth}.

\begin{figure*}[]
    \centering
    \includegraphics[width=0.9\linewidth]{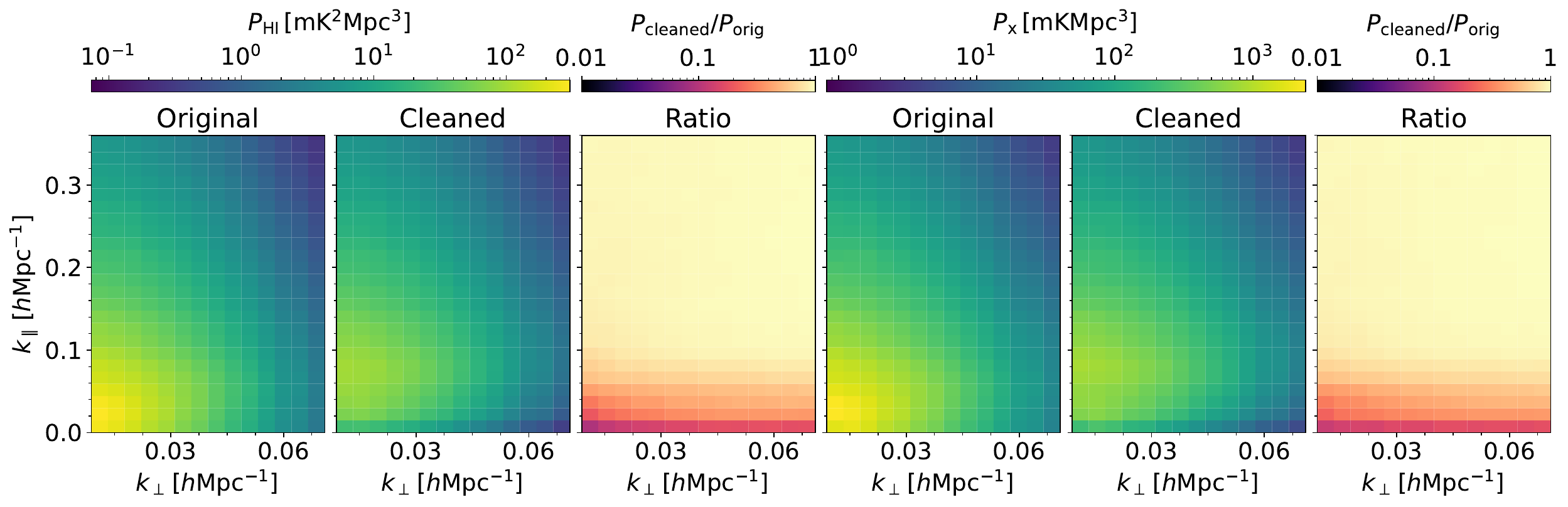}
    \includegraphics[width=0.9\linewidth]{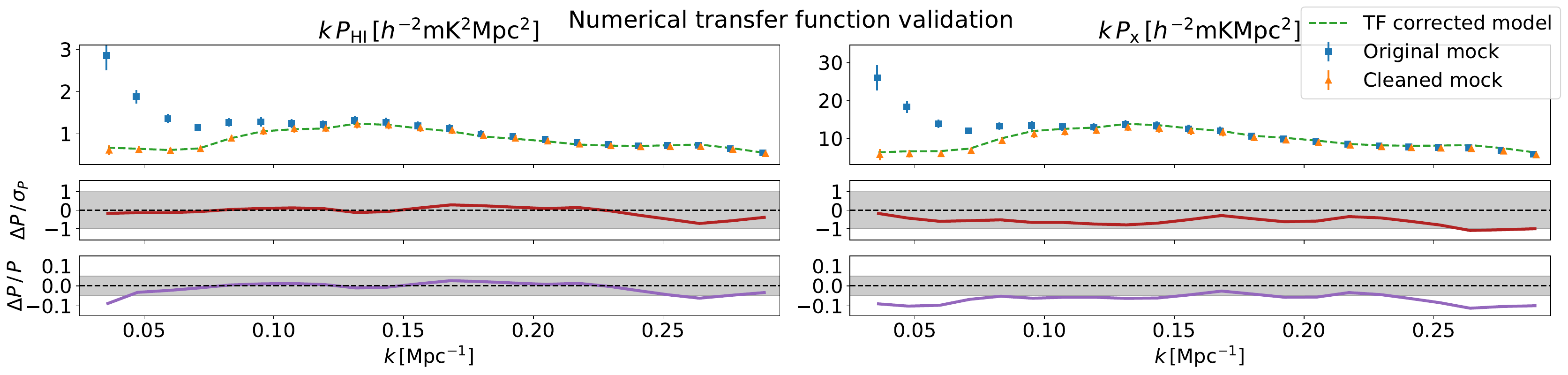}
    
    \caption{
        Validation on the accuracy of the transfer function correction from an end-to-end pipeline.
        The top panels show the cylindrical power spectra similar to \autoref{fig:estvalidation}, 
        for \hi\ auto-power and cross-power, respectively.
        For each type of the power spectrum, the left panel shows the power spectrum of the mock without foreground cleaning effects, 
        the centre panel shows the mock with foreground cleaning effects, and the right panel shows the ratio between the two.
        The bottom panels show the 1D power spectra, with the blue square representing the mock power spectrum,
        the orange triangle representing the mock with foreground cleaning effects, 
        and the green dashed line representing the input model power spectrum with numerical transfer function corrections.
        The centre of the bottom panels shows the fractional differences between the mock average and the model over the standard deviation, with the shaded area denoting the $\pm1\sigma$ region.
        The bottom row shows the fractional difference between the mock average and the model over the model, with the shaded area denoting the $\pm5\%$ region. \github{https://github.com/meerklass/meer21cm/blob/main/papers/validation/plot_02.ipynb} 
    }
    \label{fig:tfvalidation}
\end{figure*}

In the top centre panels of \autoref{fig:estvalidation}, we show the average of the galaxy auto-power against the input model.
The mock power is in good agreement $\sim 5\%$ with the input model across all scales,
except for large $|\bm{k}|$ where the measurement is close to the Nyquist frequency.
As a result of this, the cross-power, shown in the top right panels,
shows good agreements with the model at $k_\parallel \lesssim 0.2\,{h{\rm Mpc}^{-1}}$.

For cosmological analysis with MeerKLASS UHF data,
we are primarily interested in linear and BAO scales, $|\bm{k}| \txteq{<} 0.2\,{h{\rm Mpc}^{-1}}$.
Therefore, by choosing the appropriate $\bm{k}$-cuts that exclude $|\bm{k}_\perp| \txteq{>} 0.06\,{h{\rm Mpc}^{-1}}$ scales affected by the beam,
we can accurately estimate the power spectrum. Any penalty on sensitivity from these $\bm{k}$-cuts is low because the signal-to-noise is already poor in these beam-suppressed regions.
The results for the 1D power spectrum monopole are shown in the bottom panels of \autoref{fig:estvalidation}.
Across all scales $0.02\,{<}\,k\,{<}\,0.3\,{h{\rm Mpc}^{-1}}$, 
the mock power is within $1\sigma$ deviation from the model. 
The cosmological scales $k\txteq{<}0.15\,{h{\rm Mpc}^{-1}}$ are accurately recovered,
with per-cent level accuracy and deviations smaller than $0.5\sigma$ (note again we assume averaging of 10 independent patches).
We also briefly discuss and validate higher-order multipoles of the power spectrum in \appref{apdx:multipoles}.

A common feature of 21\,cm intensity mapping data is the presence of flagged frequency channels, typically arising from radio frequency interference (RFI). While mitigation strategies continue to improve, some level of flagging is unavoidable and is expected to persist in future datasets. It is therefore important to ensure that such missing or masked channels do not bias the recovered clustering signal or impact the robustness of the analysis pipeline. We provide a dedicated validation of the effect of flagged channels in \appref{apdx:flagging}, demonstrating that our methodology remains stable under these conditions.

Our results establish the consistency of our mock simulation pipeline and modelling framework,
and show that the power spectrum statistics are accurately recovered from the mock data.

\subsection{Transfer function}
\label{subsec:transferfunction}

Given the accuracy of the mock pipeline, we now add in simulated foregrounds and apply the PCA cleaning procedure to examine the accuracy of the transfer function corrections.

\begin{figure}[htbp]
    \centering
    \includegraphics[width=\linewidth]{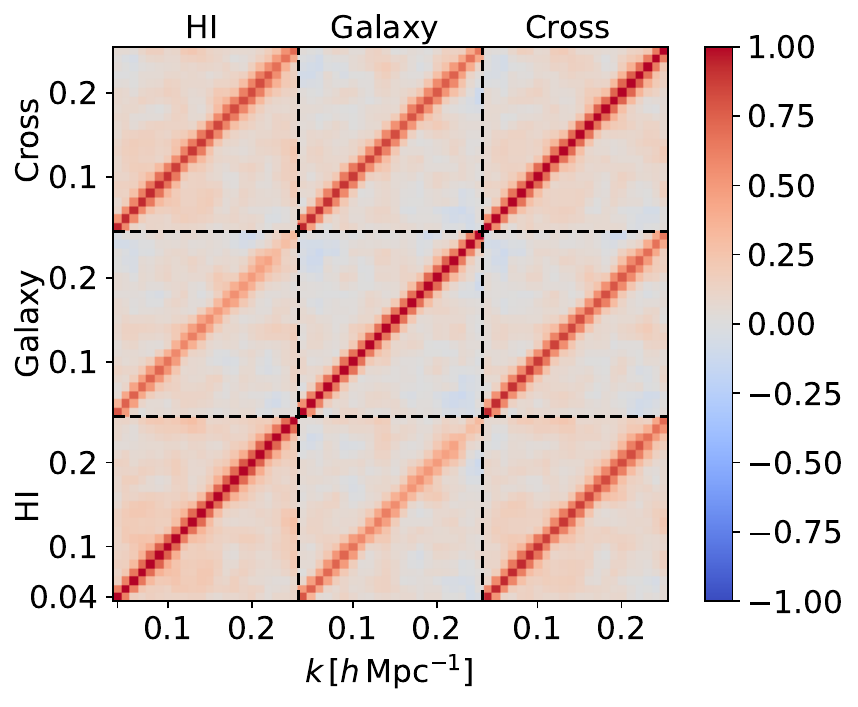}
    
    \caption{
        The correlation matrix of different $k$-bins for the simulated power spectrum data vector,
        including the \hi-auto, galaxy-auto, and cross-power.
        The correlation matrix is calculated from the 512 mock realisations.
        The \hi-auto and cross-power are the ones including foreground cleaning effects as shown in \autoref{fig:tfvalidation}.
        \github{https://github.com/meerklass/meer21cm/blob/main/papers/validation/fitting.ipynb}
    }
    \label{fig:corr}
\end{figure}

\begin{figure*}[htbp]
    \centering
%

\begin{tikzpicture}[
    class/.style={
      draw=black, fill=black!3,
      rectangle, rounded corners=3pt,
      minimum width=30mm, minimum height=6.5mm,
      align=center, font=\ttfamily\small, inner sep=2pt
    },
    class_wide/.style={
      class, minimum width=34mm
    },
    baseclass/.style={
      class, fill=black!10, font=\ttfamily\small\bfseries
    },
    helper/.style={
      draw=black!55, fill=black!1, densely dashed,
      rectangle, rounded corners=3pt,
      minimum width=28mm, minimum height=6mm,
      align=center, font=\ttfamily\small, inner sep=2pt
    },
    module/.style={
      draw=black!50, fill=black!2,
      rectangle, rounded corners=2pt,
      minimum width=32mm, minimum height=5mm,
      align=left, font=\footnotesize, inner sep=2pt
    },
    arrow/.style={->, >=stealth, thick, shorten >=1pt, shorten <=1pt},
    inherit/.style={arrow},
    used_by/.style={->, >=stealth, thick, dashed, shorten >=1pt, shorten <=1pt},
    darrow/.style={->, >=stealth, thick, dashed, gray, shorten >=1pt, shorten <=1pt},
    annotate/.style={font=\footnotesize\itshape, text=black!50},
  ]
  
  \node [helper] (Wcs) at (-5.7, 0.0)
    {\strut WcsSkyMap};
  \node [helper] (SkyMap) at (-5.7, -1.0)
    {\strut SkyMap};
  \node [helper] (Hpx) at (-5.7, -2.0)
    {\strut HealpixSkyMap};
  \draw [inherit] (SkyMap.north) -- ++(0,0.5mm) -| (Wcs.south);
  \draw [inherit] (SkyMap.south) -- ++(0,-0.5mm) -| (Hpx.north);
  
  \node [helper] (CosmoPar) at (2.64, 0.0)
    {\strut CosmologyParameters};
  
  
  \node [baseclass, class_wide] (Spec) at (-1.2, 0)
    {\textbf{Specification}};
  
  \node [class, class_wide] (CosmoCalc) at (-1.2, -2.8)
    {CosmologyCalculator};
  \node [class] (FieldPS) at (-6.0, -4.6)
    {FieldPowerSpectrum};
  
  \draw [inherit] (Spec.south) -- ++(0,-0.4) -| (CosmoCalc.north);
  \draw [inherit] (Spec.west) -- ++(-0.8,0) |- (FieldPS.east);
  
  \draw [used_by] (CosmoPar.south) -- ++(0,-0.3) -| (CosmoCalc.north east);
  
  \node [class, class_wide] (ModelPS) at (-1.2, -4.6)
    {ModelPowerSpectrum};
  \draw [inherit] (CosmoCalc.south) -- (ModelPS.north);
  
  \node [class, class_wide] (PS) at (-3.5, -6.5)
    {PowerSpectrum};
  
  \draw [inherit] (ModelPS.south) -- ++(0,-0.3) -| (PS.north east);
  \draw [inherit] (FieldPS.south) -- ++(0,-0.3) -| (PS.north west);
  
  \node [class, class_wide] (Mock) at (-3.5, -8.5)
    {MockSimulation};
  \draw [inherit] (PS.south) -- (Mock.north);
  
  \node [class, class_wide] (HIGal) at (-3.5, -10.4)
    {HIGalaxySimulation};
  \draw [inherit] (Mock.south) -- (HIGal.north);
  
  \draw [used_by] (SkyMap.east) -- ++(1.2,0) |- (Spec.south west);
  
  \node [helper] (TF) at (0.8, -10.4)
    {\strut TransferFunction};
  \draw [used_by] (Mock.east) -- ++(0.5,0) |- (TF.west);
  
  \node [helper] (FG) at (-9.0, -1.0)
    {\strut ForegroundSimulation};
  \draw [used_by] (Wcs.south west) -- ++(0.0,0.0) -| (FG.north);
  
  \node [helper] (Infer) at (-7.8, -8.5)
    {\strut Inference};
  \draw [used_by] (PS.west) -- ++(-0.8,0) |- (Infer.east);
  
  
  \node [draw=black!40, fill=black!1, rectangle, rounded corners=4pt,
         minimum width=38mm, minimum height=60mm,
         anchor=north west,
         label={[shift={(0,-0.5mm)}]north:{\footnotesize\bfseries Supporting modules}}]
    (modbox) at (2.8, -1.4) {};
  
  \node [module] at ([yshift=-5mm]modbox.north)  {\texttt{grid} --- 3D gridding};
  \node [module] at ([yshift=-15mm]modbox.north) {\texttt{io} --- map I/O};
  \node [module] at ([yshift=-25mm]modbox.north) {\texttt{telescope} --- presets};
  \node [module] at ([yshift=-35mm]modbox.north) {\texttt{stack} --- source stacking};
  \node [module] at ([yshift=-45mm]modbox.north) {\texttt{plot} --- visualisation};
  \node [module] at ([yshift=-55mm]modbox.north) {\texttt{util} --- miscellaneous};
  
  \end{tikzpicture}
  
    \caption{Overview of the \meer\ code structure.
      The main classes are denoted with solid boxes,
      and utility classes are denoted with dashed boxes.
      Solid arrows denote class inheritance, and dashed arrows
      denote that the class at the start of the arrow is used by the class at the end of the arrow
      through attributes or methods.
      Modules on the right provide supporting functionality.}
    \label{fig:code_structure}
  \end{figure*}
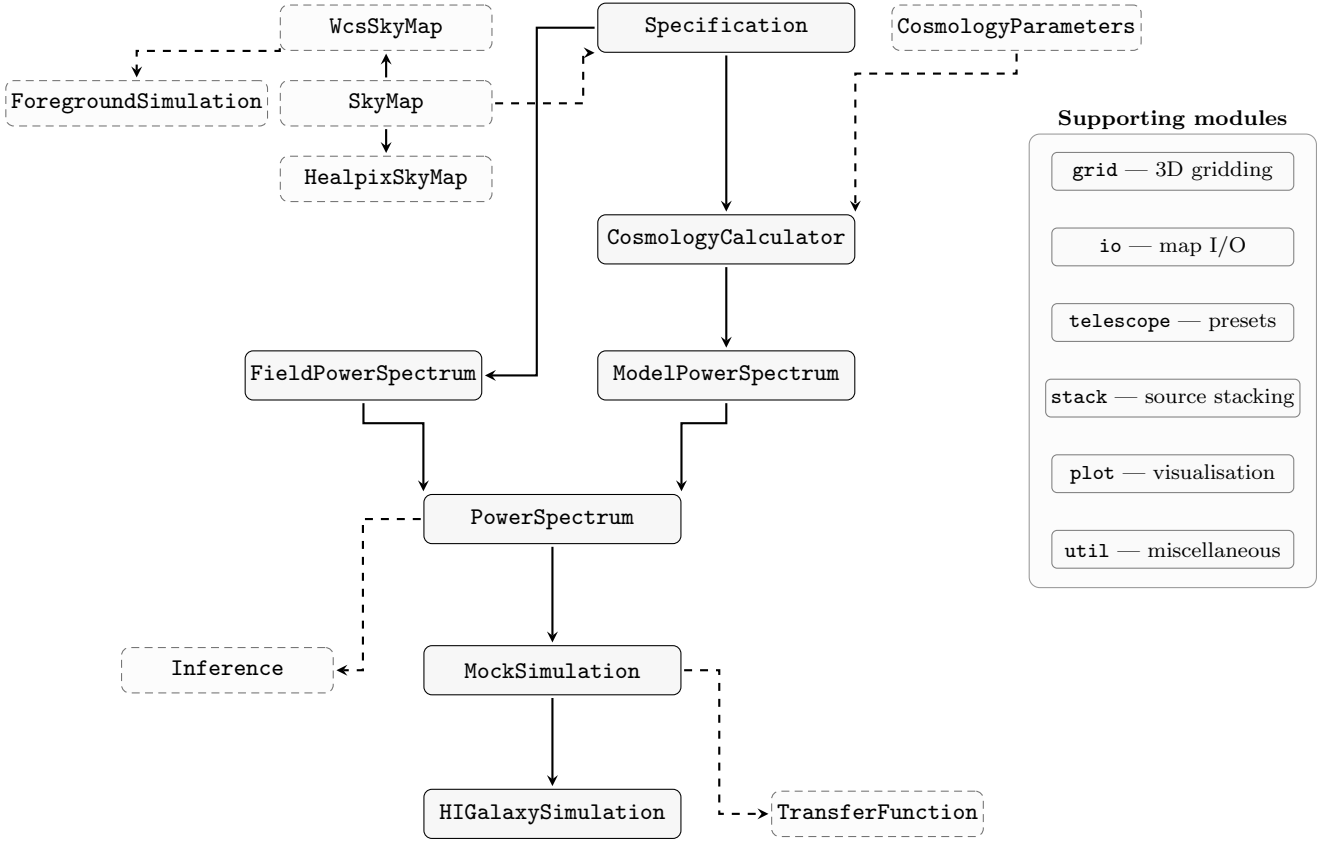

As described in \secref{subsec:TFcorrection}, 
the transfer function correction is calculated by injecting the mock \hi\ signal
into the total map data, and then calculating the signal loss according to \autoref{eq:TF}.
Following \cite{2023MNRAS.523.2453C}, we calculate the transfer function at the level of the 1D power spectrum.
For simplicity, in each mock realisation, we generate another independent mock \hi\ signal to calculate
one realisation of the transfer function, and the resulting transfer function is averaged over the mock realisations.

The effects of PCA cleaning are shown in the top panels of \autoref{fig:tfvalidation}.
Comparing the original mock power spectrum with the mock power spectrum after PCA cleaning,
we see a significant decrease in power at small $k_\parallel$.
This is expected, as foreground removal filters out the signal at large frequency intervals where foregrounds are dominant.
The amplitude of the signal loss is almost entirely determined by $k_\parallel$, since PCA only operates on the frequency direction.
Comparing the signal loss between auto- and cross-power, we see that the signal loss is identical for both,
a phenomenon that is studied in \cite{2023MNRAS.523.2453C} and proved in \cite{2025MNRAS.542L...1C}.
The $k_\parallel$-only dependence of the signal loss breaks down at the smallest $|\bm{k}_\perp|$,
suggesting that the signal loss effects in wide-angle modes are more complicated to model.
We therefore exclude the $k_x\txteq{<}0.015\,{h{\rm Mpc}^{-1}},\,k_y\txteq{<}0.015\,{h{\rm Mpc}^{-1}}$ modes from the 1D binning.

After calculating the transfer function, we apply it to the model power spectrum and
compare the results with the mock as shown in the bottom panels of \autoref{fig:tfvalidation}.
Comparing the uncleaned and cleaned mock power spectra, we see the scale-dependent signal loss affecting
scales $k \txteq{\lesssim} 0.15 \,{h{\rm Mpc}^{-1}}$, affecting the linear RSD and BAO scales.
Recall that an optimistic choice of $N_{\rm fg}=5$ has been made for the mock simulation,
and our results highlight the importance of data quality to mitigate signal loss for cosmology with MeerKLASS UHF survey.
After applying the transfer function corrections,
the model power spectra agrees with the cleaned mock within $0.5\,\sigma$ in the auto-power,
suggesting unbiased recovery of the power spectrum statistics.
The accuracy of the recovery of the cross-power is slightly poorer at $0.5\,\sigma$ to 
$1\,\sigma$ level, which is primarily due to the effects of the different weighting
for the auto- and cross-power.
For the galaxy density field, a different set of survey geometry and redshift distribution
is used, which is not reflected in the transfer function correction, since
the calculation is performed by cross-correlating the mock \hi\ map with an injected cleaned map.
Nevertheless, the accuracy is within $1\,\sigma$ level and will be sufficient for current data analysis requirements.

We further emphasise that PCA cleaning induces an additional window function
that mixes different $\bm{k}$-scales, which is not considered in the current modelling.
In principle, the PCA effects can be analytically modelled, and we present a preliminary
effort in \autoref{apdx:pcawindow}.

For reference, in \autoref{fig:corr}, we show the correlation matrix of the power spectrum data vector,
including the \hi-auto, galaxy-auto, and cross-power.
The correlation matrix shows clear correlations between nearby $k$-bins,
indicating the effects from the survey geometry.
An illustration of parameter inference using the mock data and covariance is shown in \appref{apdx:parameterinference}.

\section{Code structure and example use cases}
\label{sec:code}
In this section, we briefly describe the code structure of \meer,
and provide a few simple use cases that demonstrate the power and flexibility of the package.
A schematic diagram of the code structure is shown in \autoref{fig:code_structure}.

As discussed in \secref{sec:intro}, the \meer\ pipeline is designed to closely follow
survey specifications first and foremost, instead of starting from simulation and modelling in cubic boxes\footnote{If subsequent data analysis is not needed, simulations and power spectrum calculations in custom cubic boxes can be done. See \href{https://meer21cm.readthedocs.io/en/latest/cookbook/mode_mixing.html}{this example}.}.
Therefore, the base class for the entire codebase is the \pyth{Specification} class,
which includes the survey area, frequency range, telescope beam, and survey geometry such as the number of
hit counts at each pixel of each frequency channel.
The \pyth{Specification} class can then be used to read in survey data,
perform map-level processing such as map smoothing, trimming, and PCA cleaning.
For example, to read the MeerKLASS 2019 L-band data\footnote{Publicly available at \href{https://meerklass.org/HIdata.html}{meerklass.org/HIdata.html}} \citep{2021MNRAS.505.3698W},
we can use the following code snippet:
\begin{python}
from meer21cm import Specification
from meer21cm.plot import plot_map

map_file = '../2019/MK_2019_maps.pkl'
sp = Specification(
    pickle_file=map_file,
    survey='meerklass_2019',
    band='L',
    ra_range=[153,172],
    dec_range=[0.5,6.6],
)
sp.read_from_pickle()
plot_map(sp.data, sp.wproj, W=sp.W_HI)

\end{python}
\begin{figure}[H]
    \centering
    \includegraphics[width=0.79\linewidth]{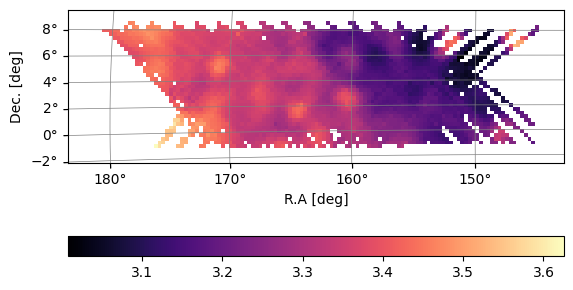}
\end{figure}

For a given survey dataset, we need input fiducial cosmology to perform
power spectrum estimation.
The cosmological model is defined by the \pyth{CosmologyParameters}\newline class,
which also handles the calculation of background quantities through \textsc{astropy} \citep{2022ApJ...935..167A}, 
as well as the matter power spectrum,
through either the Boltzmann solver code \textsc{camb} \citep{2000ApJ...538..473L} or the emulator \textsc{baccoemu} \citep{2023MNRAS.520.3725P}.
It is then used by the \pyth{CosmologyCalculator} class, which inherits from the \pyth{Specification} class to be consistent with survey specifications.
The \pyth{CosmologyCalculator} class can then be used to calculate the matter power spectrum,
survey volume, beam size in comoving distances and other quantities under fiducial cosmology.

Inherited from the \pyth{CosmologyCalculator} class, the \pyth{ModelPowerSpectrum}
class handles the calculation of the tracer model power spectrum,
which additionally includes the tracer dependant quantities and redshift space distortions on top of the matter power.
It includes two different sets of tracer settings,
allowing calculation of the auto- and cross-power spectrum for different tracers.
The \pyth{FieldPowerSpectrum} class, on the other hand,
handles power spectrum estimation for the tracer fields.
Given the estimation grid and weights for the tracer fields,
it calculates the power spectrum estimator and associated renormalisation.
It also supports the auto-power and cross-power calculation for different tracers.

The \pyth{ModelPowerSpectrum} and \pyth{FieldPowerSpectrum} classes are not meant
to be used directly, but rather to be inherited by the \pyth{PowerSpectrum} class.
Through the dependencies, the \pyth{PowerSpectrum} class can
calculate the model power spectrum and the estimator,
with the observational effects automatically applied based on the survey specifications.
The \pyth{PowerSpectrum} class provides an interface to
grid the sky map, as well as optionally the galaxy catalogue,
to Cartesian estimation grids. The gridded fields are then used for power spectrum estimation.
Through calculating the grids and weights, the observational effects are
propagated to the models under the hood.
For example, following the read in of the survey data shown in the previous example\footnote{The \pyth{Specification} class needs to be replaced by the \pyth{PowerSpectrum} class in the previous example code snippet.},
we can calculate the box dimensions corresponding to the survey volume
and the 1D model power spectrum as follows:
\begin{python}
import numpy as np
import matplotlib.pyplot as plt
# define model power spectrum
sp.k1dbins = np.linspace(0.01,1.5,21)
sp.tracer_bias_1 = 1.0
sp.omega_hi = 5e-4
sp.mean_amp_1 = "average_hi_temp"
sp.get_enclosing_box()
p1d_noobs,_,_ = sp.get_1d_power(
    sp.auto_power_tracer_1_model
)

# specify observational effects
sp.sigma_beam_ch = 0.4
sp.include_beam = [True,False]
sp.compensate = [True,True]
p1d_obs,keff,_ = sp.get_1d_power(
    sp.auto_power_tracer_1_model
)

plt.plot(
    keff,
    p1d_obs*1e6,
    label='With obs'
)
plt.plot(
    keff,
    p1d_noobs*1e6, 
    label='Without obs', 
    ls='--'
)
plt.yscale('log')
plt.xlabel(r'$k\,[{\rm Mpc}^{-1}]$')
plt.ylabel(r'$P(k)\,[{\rm mK}^2 {\rm Mpc}^3]$')
plt.legend()
\end{python}
\begin{figure}[H]
    \centering
    \includegraphics[width=\linewidth]{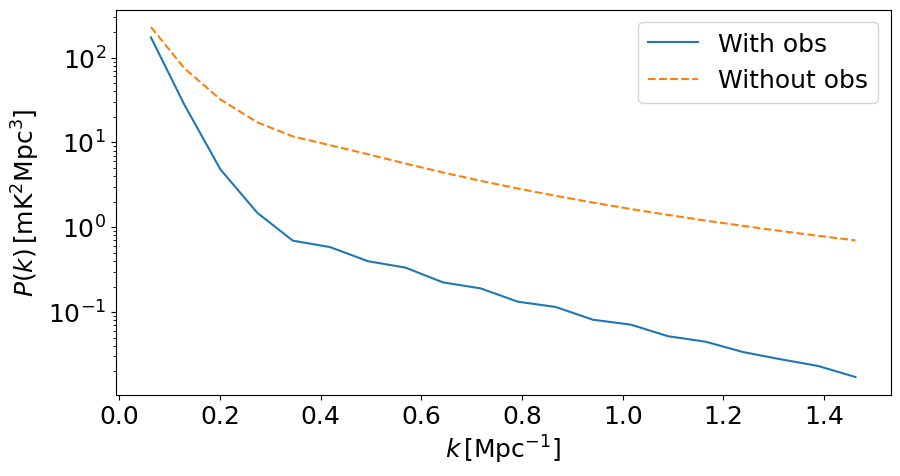}
\end{figure}

We note that, in the above example, the calculation of the model 1D power spectrum
has taken the survey volume into account, where
\pyth{sp.auto_power_tracer_1_model} is calculated on the 3D $\bm{k}$-grids.
After specifying the inclusion of the observational effects to \pyth{sp}, in this case beam attenuation and gridding compensation, the 3D power spectrum is automatically
updated under the hood, which gives the differences we see in the code output.
Specific modelling choices and inclusion of observational effects
can be controlled through the attributes of the \pyth{PowerSpectrum} class.
Therefore, by passing the attributes of the \pyth{PowerSpectrum} class
as inputs, we can use the \pyth{meer21cm.inference} module to perform parameter inference.
An example is shown in \autoref{fig:posterior} and the attached link therein.

Through reading in the survey data and specifying the model and the analysis choices,
the \pyth{PowerSpectrum} class carries the necessary information to
generate mock observations that match the survey data.
Therefore, the \pyth{MockSimulation} class inherits from the \pyth{PowerSpectrum} class,
and provides an intuitive routine to generate mock observations.
For example, we can calculate the survey volume box dimensions for the MeerKLASS L-band deep field data\footnote{Also publicly available at \href{https://meerklass.org/HIdata.html}{meerklass.org/HIdata.html}} \citep{2025MNRAS.537.3632M},
and generate the mock \hi\ field:
\begin{python}
from meer21cm import MockSimulation
from meer21cm.plot import plot_map
import matplotlib.pyplot as plt

mock = MockSimulation(
    seed=12345,
    # pre-defined survey specifications
    survey='meerklass_2021',
    band='L',
    # specify field 1 to have temp unit
    mean_amp_1 = 'average_hi_temp',
    omega_hi = 5e-4,
    tracer_bias_1 = 1.5,
)

# visualize the mock hi field
plt.imshow(
    mock.mock_tracer_field_1.mean(-1)*1e3,
    origin='lower',
    extent=[
        0,
        mock.box_len[1],
        0,
        mock.box_len[0],
    ],
    cmap='magma',
)
cbar = plt.colorbar(shrink=0.7)
cbar.set_label('[mK]')
plt.xlabel('y [Mpc]')
plt.ylabel('x [Mpc]')

# propagate the mock hi field to sky map
mock.data = mock.propagate_mock_field_to_data(
    mock.mock_tracer_field_1
)
plot_map(mock.data * 1e3,mock.wproj,W=mock.W_HI)
\end{python}
\begin{figure}[H]
    \centering
    \includegraphics[width=\linewidth]{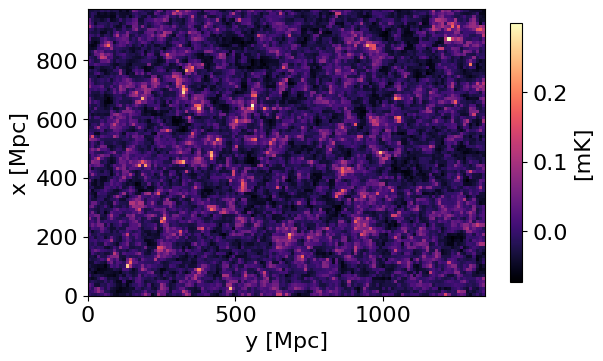}
\end{figure}
\begin{figure}[H]
    \centering
    \includegraphics[width=\linewidth]{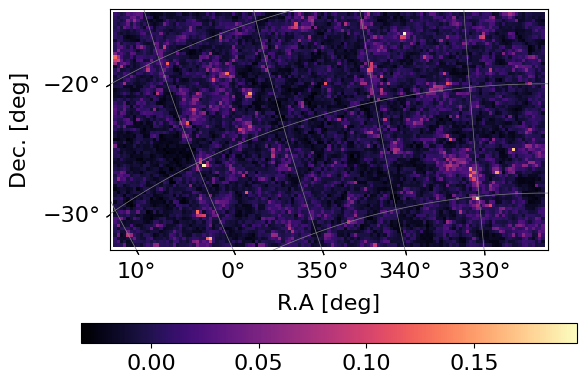}
\end{figure}
In the above example,
\meer\ calculates the box dimension and mock \hi\ field under the hood,
so that the user can directly invoke the \pyth{mock.mock_tracer_field_1}
to retrieve the mock \hi\ field.
The mock field can then be gridded to the sky map cube, which can then
be used to construct mock observations, perform signal injection tests and more.
The signal injection tests are particularly useful in the data analysis,
for the calculation of the transfer function as described in \secref{subsec:TFcorrection}.
By passing the attributes of the \pyth{MockSimulation} class as inputs,
the \pyth{TransferFunction} class automates the calculation of the transfer function.
The accuracy of the calculation has been demonstrated in \secref{subsec:transferfunction},
with the code example attached in the link following \autoref{fig:tfvalidation}.

While it is not discussed in this work, \meer\ has also been used to perform
emission line stacking analysis to produce the results reported in \cite{2025MNRAS.537.3632M} and \cite{2025ApJS..279...19C}.
As part of the analysis, the \meer\ pipeline supports simulations of the 21\,cm emission line profiles of
individual galaxies, instead of directly generating the 21\,cm brightness temperature field.
The simulation can be performed by the \pyth{HIGalaxySimulation} class,
and we refer the reader to \cite{2025ApJS..279...19C} for more details on the simulation procedure.
Both the simulated and the observed intensity maps can be used to perform stacking analysis,
with the functions provided in the \pyth{meer21cm.stack} module.

\section{Future development}
\label{sec:future}

\meer\ is intended to remain a continually evolving, community-driven toolkit, with its structure and design explicitly chosen to support long‑term development. 
The following points outline the key areas in which we anticipate active growth and future enhancement.

\begin{itemize}
    \item \textbf{Plane-parallel departure and multipole estimator}.
    For current MeerKLASS data analysis, where patches of a few hundred square degrees are used,
    the plane-parallel is an accurate approximation.
    However, for future data analysis where inter-patch calibration is performed and patches are combined into one survey lightcone,
    the increased survey volume will require dedicated simulations beyond plane-parallel,
    together with accurate estimators of the power spectrum multipoles (e.g. \citealt{2006PASJ...58...93Y}).
    The calculation of power spectrum multipoles is performed on the Cartesian grid currently, and can be extended to a spherical harmonic basis in the future \citep{2017JCAP...07..002H}, or Spherical Fourier-Bessel bases \citep{2016ApJ...833..242L,2024PhRvD.110h3525W}.
    \item \textbf{Analytical covariance estimation}.
    While the current \meer\ pipeline can robustly estimate the covariance of the measurements through mock simulations,
    the underlying model of \hi\ and \hi-galaxy cross-correlation is not clear due to lack of current measurements.
    Therefore, we need to develop an analytical framework to be implemented into \meer, to compare against 
    mock-based covariance estimation for validation (e.g. \citealt{2025JCAP...01..145R}).
    The primary focus of the modelling effort will be to include the survey geometry, the effects of the beam smoothing
    and PCA cleaning beyond the flat-sky approximation in both power spectrum and covariance estimation.
    Higher-order corrections, such as the super-sample covariance \citep{2013PhRvD..87l3504T}, should also be included in the pipeline.
    \item \textbf{Cosmological parameter inference}.
    In \appref{apdx:parameterinference}, we have demonstrated that the \meer\ pipeline can perform unbiased parameter inference.
    However, in this paper we assume fixed cosmology where the constraining power of the power spectra is
    used entirely to constrain tracer properties.
    In future work, we will investigate further into constraining cosmological parameters
    using the \meer\ pipeline.
    The primary focus will be to validate the recovery of the AP shifts assuming different fiducial and true cosmology.
    More advanced modelling of the tracer power spectrum will be integrated into \meer,
    such as the effective field theory of the large scale structure (e.g. \citealt{2012JHEP...09..082C, 2014PhRvD..89d3521H}),
    in particular for multi-tracer formalism \citep{2022JCAP...04..021M,2024MNRAS.532..783Z}.
    This will be achieved through interfacing with existing software packages, such as \textsc{pbj} \citep{2023JCAP...12..025M}.
    Furthermore, we will investigate the potential prior volume effects
    commonly seen in LSS full-shape analysis (e.g. \citealt{2023JCAP...01..028C, 2026JCAP...04..030T}).
    Combination of the intensity mapping likelihoods with external likelihoods will be implemented,
    through interfacing with \textsc{cobaya} \citep{2021JCAP...05..057T}.
    \item \textbf{High-fidelity mock}.
    The current \meer\ pipeline uses a lognormal simulation routine to generate the mock \hi\ field and the galaxy positions.
    It serves as an efficient tool for validating the power spectrum estimation pipeline and performing numerical transfer function calculations.
    However, as we go beyond the current approximations,
    we need to develop high-fidelity mock products to enable the validation of more advanced modelling choices.
    \item \textbf{Implementation of various foreground cleaning methods}.
    We have discussed extensively the corrections for foreground removal effects using PCA in \secref{subsec:transferfunction} and \appref{apdx:pcawindow}. Given the complexity in the data analysis, alternative methods such as Gaussian Process Regression (e.g. \citealt{2022MNRAS.510.5872S}), generalized morphological component analysis (e.g. \citealt{2020MNRAS.499..304C}) and mPCA (e.g. \citealt{2025A&A...703A.222C}) may be useful in systematics mitigation. Validating the impact of these methods on the power spectrum estimation will be important for incorporating them into the \meer\ pipeline in the future.
    \item \textbf{Improved computational efficiency}.
    The current \meer\ pipeline is efficient for the current MeerKLASS data analysis.
    Nevertheless, the increased data volume for future surveys poses challenges,
    especially for the mock simulation routines.
    Moreover, the parallelisation of the pipeline currently happens
    outside the codebase in individual simulation scripts.
    Further improvements to the computational efficiency will be investigated,
    such as a dedicated worker class to allow more automated interaction between the pipeline
    and parallel processes, such as \textsc{mpi}.
    \item \textbf{Marginalisation over the residual systematic effects}.
    In this work, we focus on the post-calibration aspect of intensity mapping data analysis.
    However, residual systematic effects will be present in the actual data,
    such as beam chromaticity \citep{2021MNRAS.506.5075M}, 
    residual $1/f$ noise \citep{2021MNRAS.501.4344L,2024MNRAS.527.4717I},
    and gain non-linearity effects.
    Modelling systematics with additional nuisance parameters allows for the marginalisation over the residual systematic effects,
    providing a more robust and unbiased estimation of cosmology.
\end{itemize}

\section{Conclusion}
\label{sec:conclusion}
In this paper, we present the \meer\ pipeline, a comprehensive toolkit for \hi\ intensity mapping data analysis.
Focusing on the MeerKLASS survey, we design the \meer\ package to be
survey and data analysis oriented, with an intuitive structure for interactive use.

The \meer\ pipeline implements a detailed modelling of the \hi\ signal as well as the galaxy clustering
and the cross-correlation between the two, focusing on the various observational effects.
In particular, we structure the computation of the model power spectrum to follow exactly the survey specifications.
The modelling always happens on the level of the 3D $\bm{k}$-space, where the 1D multipole power spectra are calculated
through the averaging of the 3D power spectrum instead of a theoretical integral of $\mu$-space.
This ensures that the modelling takes into account the exact survey volume, the coarse $\bm{k}$-sampling and the
highly non-uniform resolution of \hi\ intensity mapping surveys.
We detail the estimator for all three power spectra cases (auto-\hi, auto-galaxy and cross-correlation),
with the description of the weight renormalisation differing between the three.
The estimator and the modelling description are consistent in the normalisation and the modelling of
instrument effects and survey window functions.
The formalism, as laid out in \secref{sec:formalism}, serves as the reference for future MeerKLASS UHF data analysis,
and will be further developed as discussed in \secref{sec:future}.

The modelling of the power spectra in the \meer\ pipeline requires the specification of the survey and the telescope,
the properties of the tracers, the cosmological model, 
the choices of data analysis, such as the number of foreground modes to be removed,
and the weight assignment scheme when gridding to the Cartesian coordinates.
Following the structure described in \secref{sec:code}, the \meer\ pipeline
consistently incorporates these choices into the calculation of the power spectra,
as well as the calculation of the transfer function through mock simulation routines.
The consistency and the accuracy of the modelling have been demonstrated in \secref{sec:validation}.
First, we validate the accuracy of the lognormal simulation of the \hi\ density field and the galaxy positions.
By directly comparing the input model power spectra against the simulated tracer density fields,
we find that the lognormal simulation is accurate to within 1\% of the input model power spectra,
as demonstrated in \autoref{fig:mockvalidation}.
Then, we pass the simulated tracers to the mock observation pipeline,
to generate the observed intensity maps and galaxy catalogues, which
are then re-gridded to the Cartesian coordinates to perform the power spectrum estimation.
The mock observation follows a typical redshift sub-band for a single patch of the MeerKLASS UHF survey,
with the survey area of $\rm {\sim}\,750\,deg^2$ and the redshift sub-band of $0.6\,{<}\,z\,{<}\,0.8$.
The results are shown in \autoref{fig:estvalidation}.
We find that, with the observation effects included,
the deviations from the mock power spectra are $\sim \pm 1\%$ of the model power spectra
for $k < 0.2\,{h{\rm Mpc}^{-1}}$ in the case of the galaxy auto-power.
The accuracy decreases slightly for the cross-power and the \hi\ auto-power,
which is primarily due to the effects of the beam smoothing that requires modelling
beyond the flat-sky and plane-parallel approximations.
Still, the results are within the $\pm 5\%$ accuracy of the model,
and the deviations are smaller than $0.5 \sigma$ of the signal variance assuming a total of 10 patches.

We further test the accuracy of the corrections for the foreground cleaning effects
in the \meer\ pipeline.
By running the numerical transfer function calculations and applying the corrections
to the model power spectra as shown in \autoref{fig:tfvalidation}, we find that the foreground cleaned mock power spectra
agree with the model power spectra within $\pm 5\%$ accuracy for
$0.05\,{h{\rm Mpc}^{-1}} \txteq{<} k \txteq{<} 0.2\,{h{\rm Mpc}^{-1}}$ in the case of the \hi\ auto-power.
The accuracy decreases for large scales $k \txteq{<} 0.05\,{h{\rm Mpc}^{-1}}$,
due to the large signal loss and wide-angle effects at these scales.
The accuracy of the cross-power, while showing good agreement with the model within $1\sigma$ of the signal variance,
shows a systematic offset of $\sim 5\%$.
This is primarily due to the different weighting scheme for the auto- and cross-power.
Additionally, we examine the possibility of applying analytical corrections of the foreground cleaning effects
in \appref{apdx:pcawindow}, where we find similar performance as the numerical corrections at $k \txteq{>} 0.05\,{h{\rm Mpc}^{-1}}$.
The validation results demonstrate the accuracy and the robustness of \meer, ensuring its suitability
for the data analysis of the MeerKLASS UHF survey.
The simulation pipeline used for the validation also serves as the basis for
generating the mock realisations needed for covariance estimation for using the multi-tracer power spectra as shown in \autoref{fig:corr}.
A preliminary example of parameter inference using the mock data and covariance is shown in \appref{apdx:parameterinference}.

The \meer\ package is designed to be easy to use and highly flexible,
with the structure and design aimed for interactive use and extensibility.
The code is open source with publicly available documentation and examples.
We expect the \meer\ package to be a valuable tool for the
MeerKLASS collaboration as well as the wider \hi\ intensity mapping community.
The continued development of the \meer\ package will be driven by the needs of the community
and the future MeerKLASS UHF data analysis.
In the future, we envision the extensive use of the \meer\ package
will allow for robust cosmological analysis of the MeerKLASS UHF survey data,
in preparation for the era of radio cosmology with the future SKAO.

\section*{Acknowledgements}

ZC and AP are funded by a UKRI Future Leaders Fellowship
[grant MR/X005399/1; PI: Alkistis Pourtsidou].
SCu acknowledges support from the UKRI Stephen Hawking Fellowship (grant reference EP/U536751/1).
JLB acknowledges funding from the project UC-LIME (PID2022-140670NA-I00), financed by MCIN/AEI/ 10.13039/501100011033/FEDER, UE.
SCa acknowledges support from the Italian Ministry of University and Research (\textsc{mur}), PRIN 2022 `EXSKALIBUR – Euclid-Cross-SKA: Likelihood Inference Building for Universe's Research', Grant No.\ 20222BBYB9, CUP D53D2300252 0006, from the Italian Ministry of Foreign Affairs and International
Cooperation (\textsc{maeci}), Grant No.\ ZA23GR03, and from the European Union -- Next Generation EU.
IPC is supported by the European Union within the Next Generation EU programme [PNRR-4-2-1.2 project No.\ SOE\textunderscore0000136, RadioGaGa]. 
JW acknowledges support from the National Natural Science Foundation of China (NSFC, Grant No.\ 12573112).
DT acknowledges support from the Science and Technology Facilities Council (STFC) studentship. 
JF thanks the support of FCT - Fundação para a Ciência e a Tecnologia through national funds by these grants: UID/04434/2025 (DOI 10.54499/UID/04434/2025) and 2023.15069.PEX and through the Scientific Employment Incentive program (reference 2020.02633.CEECIND/CP1631/CT0002).

This work used data products obtained using the MeerKAT telescope, operated by the South African Radio Astronomy Observatory, which is a facility of the National Research Foundation, an agency of the Department of Science and Innovation. Specifically we used data from projects SCI-20180330-MS-01, SCI-20190418-MS-01 and SCI-20210212-MS-01.
We acknowledge the use of the Ilifu cloud computing facility, through
the Inter-University Institute for Data Intensive Astronomy (IDIA)
\clearpage
\appendix

\section{Galaxy power spectrum}
\label{apdx:galaxyps}
In this section, we give a brief derivation of the weight renormalisation for the galaxy power spectrum.
Ignoring the gridding effects, the observed galaxy overdensity field is given by
\begin{equation}
    \delta_{\rm n}^{\rm obs}(\bm{x})+1 = \frac{w^g_{\rm n}(\bm{x})\,{\rm n}_{\rm obs}(\bm{x})}{\big\langle w^g_{\rm n}(\bm{x})\,{\rm n}_{\rm obs} (\bm{x}) \big\rangle_{\rm V}} = \frac{W_{\rm gal}(\bm{x}) \,{\rm n}_{\rm gal}(\bm{x})}{\big\langle W_{\rm gal}(\bm{x}) \,{\rm n}_{\rm gal}(\bm{x}) \big\rangle_{\rm V}} = \frac{W_{\rm gal}(\bm{x})}{\big\langle W_{\rm gal}(\bm{x}) \big\rangle_{\rm V}} \big(\delta_{\rm n}(\bm{x})+1\big),
\end{equation}
where $W_{\rm gal}(\bm{x}) = w_{\rm n}^f(\bm{x}) w_{\rm n}^g(\bm{x})$ is the total weights for the galaxy density.
We use the fact that the underlying galaxy number density field and the total weights are uncorrelated, so that $\big\langle W_{\rm gal}(\bm{x}) \,{\rm n}_{\rm gal}(\bm{x}) \big\rangle_{\rm V} = \big\langle W_{\rm gal}(\bm{x}) \big\rangle_{\rm V} \big\langle \,{\rm n}_{\rm gal}(\bm{x}) \big\rangle_{\rm V}$.
Treating the factor before $ \delta_{\rm n}(\bm{x}) $ as a normalisation factor $w'(\bm{x}) = W_{\rm gal}(\bm{x}) / \big\langle W_{\rm gal}(\bm{x}) \big\rangle_{\rm V}$, the correct renormalisation is then
\begin{equation}
    \hat{P}_{\rm gal}(\bm{k}) = \frac{{\rm V}}{\langle (w'(\bm{x}))^2 \rangle_{\rm V}} \big|\tilde{\delta}_{\rm n}^{\rm obs}(\bm{k})\big|^2 = \frac{{\rm V} \langle W_{\rm gal}(\bm{x}) \rangle_{\rm V}^2}{\langle W_{\rm gal}(\bm{x})^2 \rangle_{\rm V}} \big|\tilde{\delta}_{\rm n}(\bm{k})\big|^2,
\end{equation}
which is equivalent to the estimator in \autoref{eq:pgest}.
We emphasise again that the derivation assumes that the gridding kernel and the grid-level weights are commutable,
so that the weights can be combined into a single total weight.
The derivation can then be trivially generalised to the case of cross-power to give \autoref{eq:pcest_cross}.

The shot noise modelling, on the other hand, requires a separate treatment by discretising the galaxy number density field. 
The base shot noise can be derived by discretising the galaxy number density field so that
\begin{equation}
    {\rm n}_{\rm obs}(\bm{x}) =\frac{\sum_{\rm i}\delta^3_{\rm D} (\bm{x}-\bm{x}_i)}{{\rm V}},
\end{equation}
where $\delta^3_{\rm D}$ is the 3D Dirac delta function, and $i$ iterates over all observed galaxies.
The observed galaxy overdensity field is then
\begin{equation}
    \delta_{\rm n}^{\rm obs}(\bm{x}) + 1 = \frac{w_{\rm n}^g(\bm{x})\sum_i \delta^3_{\rm D} (\bm{x}-\bm{x}_i)}{{\rm V}} \frac{\rm V}{\sum_i w_{\rm n}^g(\bm{x}_i)} 
     = \frac{w_{\rm n}^g(\bm{x})}{\sum_i w_{\rm n}^g(\bm{x}_i)} \sum_i \delta^3_{\rm D} (\bm{x}-\bm{x}_i).
\end{equation}
The power spectrum is the Fourier transform of the two-point correlation function,
\begin{equation}
\begin{split}
    P_{\rm n}(\bm{k}) = & \int {\rm d}^3 \bm{s} \langle \delta_{\rm n}^{\rm obs}(\bm{x}) \delta_{\rm n}^{\rm obs}(\bm{x}+\bm{s}) \rangle \, {\rm exp}\big[{-i\bm{k}\cdot\bm{s}}\big]
    \\ = & \int \frac{{\rm d}^3 \bm{x}}{V} {\rm d}^3 \bm{s} \, \delta_{\rm n}^{\rm obs}(\bm{x}) \delta_{\rm n}^{\rm obs}(\bm{x}+\bm{s}) \, {\rm exp}\big[{-i\bm{k}\cdot\bm{s}}\big]
    \\ = & \int \frac{{\rm d}^3 \bm{x}}{V} {\rm d}^3 \bm{s} \, \Big(\frac{w_{\rm n}^g(\bm{x})}{\sum_m w_{\rm n}^g(\bm{x}_m)}\Big)^2 \sum_{i,j} \delta^3_{\rm D} (\bm{x}-\bm{x}_i) \delta^3_{\rm D} (\bm{x}+\bm{s}-\bm{x}_j) \, {\rm exp}\big[{-i\bm{k}\cdot\bm{s}}\big].
\end{split}
\end{equation}
The shot noise is the self-pair part of the two-point correlation function, i.e. $i=j$.
It can then be computed as
\begin{equation}
\begin{split}
    P_{\rm SN}(\bm{k}) = & \int \frac{{\rm d}^3 \bm{x}}{V} {\rm d}^3 \bm{s} \, \Big(\frac{w_{\rm n}^g(\bm{x})}{\sum_m w_{\rm n}^g(\bm{x}_m)}\Big)^2 \sum_{i} \delta^3_{\rm D} (\bm{x}-\bm{x}_i) \delta^3_{\rm D} (\bm{x}+\bm{s}-\bm{x}_i) \, {\rm exp}\big[{-i\bm{k}\cdot\bm{s}}\big]
    \\ = & \int {\rm d}^3 \bm{x} \, \Big(\frac{w_{\rm n}^g(\bm{x})}{\sum_m w_{\rm n}^g(\bm{x}_m)}\Big)^2 \sum_{i} \delta^3_{\rm D} (\bm{x}-\bm{x}_i) \, {\rm exp}\big[{-i\bm{k}\cdot(\bm{x}_i-\bm{x})}\big]
    \\ = & {\rm V}\sum_i  \Big(\frac{w_{\rm n}^g(\bm{x}_i)}{\sum_m w_{\rm n}^g(\bm{x}_m)}\Big)^2 
    \\ = & {\rm V} \frac{\sum_i (w_{\rm n}^g(\bm{x}_i))^2}{(\sum_i w_{\rm n}^g(\bm{x}_i))^2} = \frac{1}{\langle n_{\rm obs} \rangle_{\rm V} } \frac{\langle w_{\rm n}^g(\bm{x})^2 \rangle_{\rm N}}{\langle w_{\rm n}^g(\bm{x}) \rangle_{\rm N}^2},
\end{split}
\label{eq:psn_deriv}
\end{equation}
where the discrete sum is rewritten as the number density weighted average.
Finally, note that the power spectrum estimator is renormalised,
which then introduces an additional factor.
Together with the gridding compensation factor, it then matches \autoref{eq:psn_deriv} with \autoref{eq:psn}.

It is worth noting, as mentioned in \secref{sec:formalism}, that the shot noise is assumed to be Poissonian to be consistent with the log-normal simulation routine described in \appref{apdx:lognormal}.
This assumption must hold at the zeroth order, since it naturally arises from the point-source nature of the galaxy density field, as seen in the derivations above.
From a field-level perspective, higher-order corrections of shot noise can be added to describe the additional stochastic relation between the galaxy number count and the underlying matter density \citep{2023PhRvD.108h3528O}.

\section{Lognormal simulation}
\label{apdx:lognormal}
In this section, we briefly describe the lognormal simulation of the \hi\ density field and the galaxy positions.

For an input power spectrum $P(\bm{k})$ on the 3D $\bm{k}$-space, the lognormal field can be generated by first generating the underlying Gaussian field.
The power spectrum of the Gaussian field is given by \citep{2011MNRAS.416.3017B}
\begin{equation}
    P_{\rm G}(\bm{k}) = \mathcal{F}\big[1 + \ln \mathcal{F}^{-1}[P(\bm{k})]\big],
\end{equation}
where $\mathcal{F}$ denotes the Fourier transform, and $\mathcal{F}^{-1}$ denotes the inverse Fourier transform.
Based on the power spectrum $P_{\rm G}(\bm{k})$, the Gaussian field can be generated in Fourier space and then transformed back to give $\delta_{\rm G}(\bm{x})$.
The lognormal field is then
\begin{equation}
    \delta_{\rm log}(\bm{x}) = \exp(\delta_{\rm G}(\bm{x}) - \sigma_{\rm G}^2/2) - 1,
\end{equation}
where $\sigma_{\rm G}$ is the normalisation factor so that
\begin{equation}
    \sigma_{\rm G}^2 = \langle P_{\rm G}(\bm{k}) \rangle.
\end{equation}

For \hi\ temperature field, the overdensity field $\delta_{\rm \hi}(\bm{x})$ is then multiplied by the average brightness temperature $\bar{T}_{\rm \hi}$ 
to give the \hi\ temperature field $T_{\rm \hi}(\bm{x})$.
For galaxy positions, we first compute the expected galaxy number count for each cell voxel,
\begin{equation}
    \langle N_{\rm g}(\bm{x}) \rangle  = (1 + \delta_{\rm g}(\bm{x})) V_{\rm cell} \bar{n}_{\rm g}(\bm{x}),
\end{equation}
where $V_{\rm cell}$ is the volume of each cell voxel, $\bar{n}_{\rm g}(\bm{x})$ is the average galaxy number density incorporating the
galaxy redshift distribution and survey masks.
The actual number of galaxies in each cell voxel is then sampled from the Poisson distribution so that
\begin{equation}
    N_{\rm g}(\bm{x}) \sim \text{Poisson}(\langle N_{\rm g}(\bm{x}) \rangle).
\end{equation}

Note that, the above routine is straightforward for simulations without RSD.
For density field simulations with RSD, \meer\ offers two options.
When \pyth{mock.rsd_from_field =  False}, we use plane-parallel approximation to generate
directly the anisotropic input power spectrum and then the lognormal fields using the aforementioned routine.
When \pyth{mock.rsd_from_field =  True}, we generate the lognormal density field in real space first, and then use the Zel'dovich approximation
to generate the peculiar velocity field so that
\begin{equation}
    \tilde{\bm{v}}(\bm{k}) = -i \mathcal{H} f \frac{\bm{k}}{|\bm{k}|^2} \tilde{\delta}_{\rm r}^{\rm m}(\bm{k}),
\end{equation}
where $\mathcal{H}$ is the conformal Hubble parameter, $f$ is the growth rate, 
and $\tilde{\delta}_{\rm r}^{\rm m}(\bm{k})$ is the Fourier transform of the lognormal matter density field in real space.
The lognormal density field in redshift space is then
\begin{equation}
    \tilde{\delta}_{\rm s}(\bm{k}) = \tilde{\delta}_{\rm r}(\bm{k}) + i \frac{k_z \hat{z}}{\mathcal{H}} \tilde{\bm{v}}_z(\bm{k}),
\end{equation}
where $\hat{z}$ is the unit vector along the line-of-sight assuming global parallel-plane approximation.
Note that, the galaxy positions in this case cannot be directly generated from the redshift-space overdensity $\delta_{\rm s}(\bm{x})$,
because the overdensity is only lognormal in real space.
In redshift-space, the overdensity is no longer lognormal with a minimum value of -1,
and negative values of $\langle N_{\rm g}(\bm{x}) \rangle$ are not allowed in Poisson sampling.
An alternative approach is to generate the galaxy positions in real space first,
and then assign peculiar velocities to the galaxies based on the generated peculiar velocity field (e.g. \citealt{2017JCAP...10..003A}).
However, this approach does not assume the perturbative expansion of the Jacobian determinant from real space to redshift space,
and causes a potential mismatch between the galaxy positions and the linear Kaiser RSD assumed.
Since we need to simulate down to physical scales of $\sim 1\,{\rm Mpc}/h$,
we adopt the following approach.
For the generated lognormal galaxy overdensity field $\delta_{\rm g}(\bm{x})$, we first define a mask function,
so that $w_{\rm mask}(\bm{x}) = 1$ for $\delta_{\rm g}(\bm{x}) > -1$, and $w_{\rm mask}(\bm{x}) = 0$ otherwise.
The power spectrum of the masked density field has an offset compared to the original power spectrum,
\begin{equation}
    P_{\rm original}(\bm{k})= P_{\rm masked}(\bm{k})\times C_{\rm norm} = P_{\rm masked}(\bm{k}) \times \frac{\sum_i [w_{\rm mask}(\bm{x}_i)]^2 }{[\sum_i w_{\rm mask}(\bm{x}_i)]^2} 
    \times \bigg(\frac{\sum_i(1 + \delta_{\rm g}(\bm{x}_i))}{\sum_i(1 + \delta_{\rm g}(\bm{x}_i)w_{\rm mask}(\bm{x}_i))}\bigg)^2,
\end{equation}
where the first factor comes from power spectrum normalisation of the grid weights, and the second factor comes from
the mean-centering of the density field.

We can then modify the masked lognormal density field to
\begin{equation}
    \delta_{\rm g}'(\bm{x}) = (1+\delta_{\rm g}(\bm{x}))^{\sqrt{C_{\rm norm}}} - 1,
\end{equation}
so that in the limit of $\delta_{\rm g}\ll 1 $, the output power spectrum is normalised while keeping the
minimum value of $\delta_{\rm g}'(\bm{x})$ at -1.
The modified density field is then used to generate the galaxy positions in redshift space.

For the results shown in the paper, we use \pyth{mock.rsd_from_field =  False}.
We verify that the results are consistent when \pyth{mock.rsd_from_field =  True} and omit the validation results for this case for simplicity.
We note that the choice of \pyth{mock.rsd_from_field =  True} is kept to allow further improvements
of the simulation to go beyond plane-parallel approximation as discussed in \secref{sec:future}.

\section{Fiducial models for mock validation}

In \autoref{tab:model}, we summarise the fiducial models used for the mock validation.
Note that we omit the input cosmological model, which is set to be the ``Planck 2018'' cosmology \citep{2020A&A...641A...6P} as mentioned in \secref{sec:intro}.

\begin{table}[ht]
    \centering
    \begin{tabular}{c|c|c|c|c|c|c|c|c}
       Notation  & $b_{\rm \hi}$ & $b_{\rm gal}$ & $\sigma_{\rm p}^{\rm \hi}$ & $\sigma_{\rm p}^{\rm gal}$ & $\Omega_{\rm \hi}$ & $r_\times$ & $\langle \rm n_{\rm obs} \rangle_V $ & $\sigma_{\rm beam}$ \\
       \hline
       Model  & 1.0 & 1.0 & $100\,[{\rm km/s}]$ & $100\,[{\rm km/s}]$ & $5\times 10^{-4}$ & 1.0 & $2\times 10^{-4}\,[{\rm Mpc}^{-3}$] & $0.54\,(1{\rm GHz}/\nu)\, [{\rm deg}]$ \\
    \end{tabular}
    \caption{Fiducial models used for the mock validation.}
    \label{tab:model}
\end{table}

We note that, since this paper focuses on validating the accuracy of the power spectrum estimation and the mock simulation,
the specific values of the input model parameters are not of primary interest.
We choose $b_{\rm \hi} = b_{\rm gal} = 1.0$ for simplicity.
The bias parameters are set to be equal to each other due to an intrinsic shortcoming of the lognormal simulation.
Since the simulation is directly generated on the input tracer power spectra,
having different bias parameters will slightly decorrelate the two lognormal density fields,
while they share the same underlying Gaussian random seed.
See discussions in \cite{2017JCAP...10..003A} in the context of matter and galaxy cross-correlation, where the cross-correlation coefficient has $\sim 2\%$ level deviations from 1.0 at $k\sim 0.2\,{\rm Mpc}^{-1}$.
For the same reason, the lognormal simulation cannot incorporate the cross-correlation coefficient $r_\times$ in the simulation,
which is equal to 1 in the case of equal bias parameters.
We set $\Omega_{\rm \hi} = 5\times 10^{-4}$, which is broadly consistent with observed \hi\ density at $z<1$ 
(see e.g. \citealt{2019MNRAS.489.1619H} and references therein).
Finally, we set $\langle \rm n_{\rm obs} \rangle_V = 2\times 10^{-4}\,[{\rm Mpc}^{-3}]$, 
which is approximately the average galaxy number density of the LRG galaxy catalogue in the first data release of
the DESI survey \citep{2025JCAP...04..012A}.

For the primary beam model, we assume a Gaussian beam, so that the beam size is determined by the width of the Gaussian profile $\sigma_{\rm beam} = \theta_{\rm FWHM}/2\sqrt{\ln 2}$, with the fiducial values listed in \autoref{tab:model}. The corresponding beam kernel is then
\begin{equation}
    \tilde{B}(\bm{k}) = \exp \Big[ - |\bm{k}_\perp|^2 \sigma_{\rm beam}^2(\nu_{\rm eff}) d_c^2(\nu_{\rm eff})\Big],
\end{equation}
where $d_c(\nu)$ is the comoving distance at the observing frequency $\nu$, and $\nu_{\rm eff}$ is the frequency corresponding to the effective redshift of the sub-band.

For the map-making process, we assume that the values of intensity maps are simple averages of the time-ordered-data.
This leads to the expression of the map-making kernel to be a simple windowing of the nearest-grid-point mass assignment scheme,
\begin{equation}
    \tilde{G}_{\rm map}(\bm{k}) = {\rm sinc}\Big( \frac{k_xH_{\rm pix}}{2} \Big){\rm sinc}\Big( \frac{k_yH_{\rm pix}}{2} \Big){\rm sinc}\Big( \frac{k_zH_{\rm ch}}{2} \Big),
\end{equation}
where $H_{\rm pix} = d_c(\nu_{\rm eff})\delta\theta_{\rm map}$ is the map pixel resolution for an angular resolution of $\delta\theta_{\rm map}$, and $H_{\rm ch} = [d_c(\nu_{\rm min})-d_c(\nu_{\rm max})]/N_{\rm ch}$ is the channel resolution in comoving distance.

Finally, for the gridding compensation kernel, it follows
\begin{equation}
    \tilde{G}_{\rm grid}(\bm{k}) = \bigg[{\rm sinc}\Big( \frac{k_xH_{x}}{2} \Big){\rm sinc}\Big( \frac{k_yH_{y}}{2} \Big){\rm sinc}\Big( \frac{k_zH_{z}}{2} \Big)\bigg]^{p},
\end{equation}
where $H_{x,y,z}$ is the grid size along each direction for the estimation grid, and $p$ is the index determined by the mass assignment scheme. For this work, we choose $p=2$ corresponding to the CIC scheme used in the validation mocks.

\section{Beam effects}
\label{apdx:beamsmooth}
\begin{figure}[b]
    \centering
    \includegraphics[width=\linewidth]{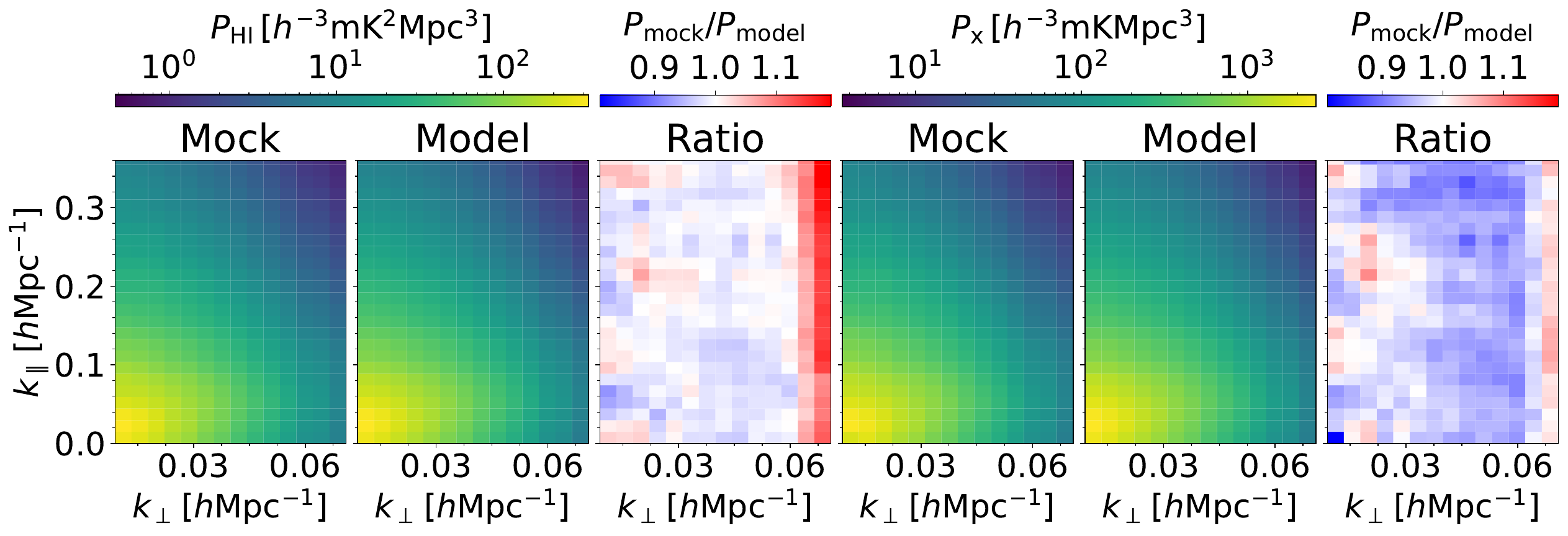}
    \includegraphics[width=\linewidth]{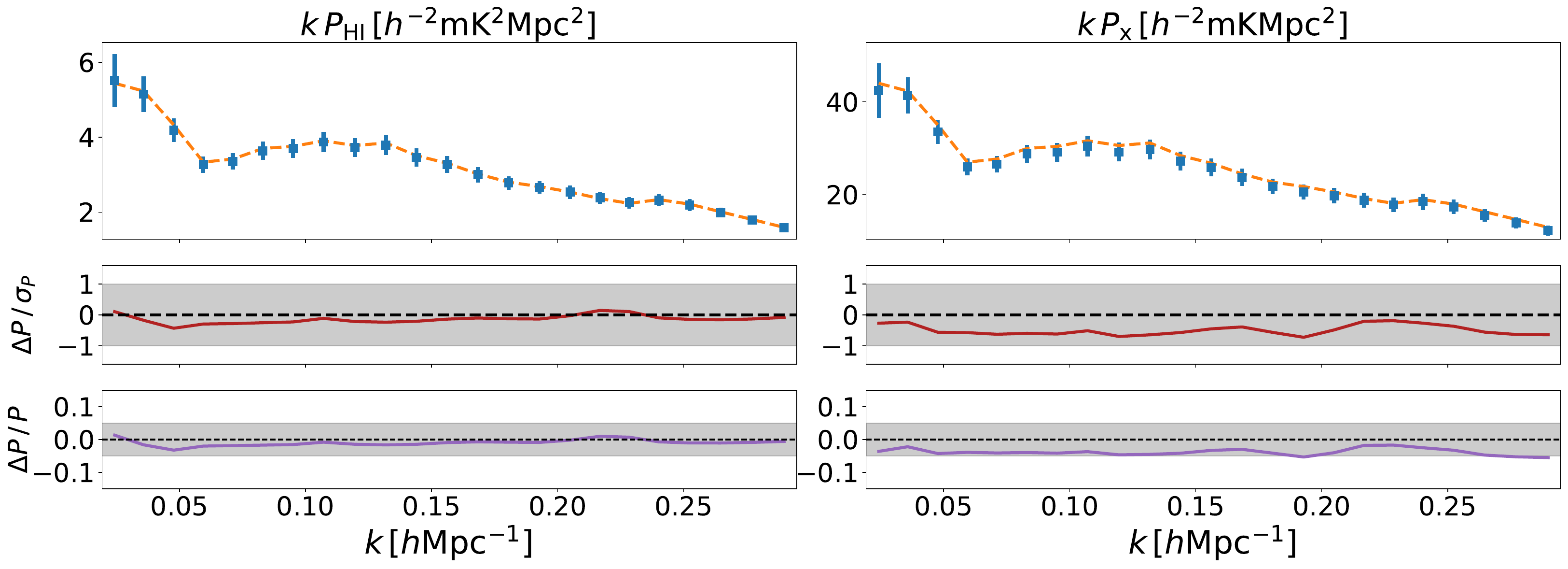}
    \caption{
        Validation of the power spectrum estimation without beam smoothing effects.
        The results shown are the same as \autoref{fig:estvalidation}, 
        but without beam smoothing effects.
        \github{https://github.com/meerklass/meer21cm/blob/main/papers/validation/plot_nobeam.ipynb}
    }
    \label{fig:nobeam}
\end{figure}

In this section, we verify the cause of the beam effect on the power spectrum estimation.
We run the pipeline without beam smoothing effects, and find that the structure at small $|\bm{k}_\perp|$, large $k_\parallel$
diminishes, as shown in \autoref{fig:nobeam}.
Comparing it with the results shown in \autoref{fig:estvalidation},
we find that the underestimation of the power spectrum at small $|\bm{k}_\perp|$, large $k_\parallel$ has disappeared.
This verifies that the beam effect is the cause of the structures we see in \secref{sec:validation}.
Excluding the beam attenuation also naturally removes the mismatch at high $|\bm{k}_\perp|$.

We emphasise that, the mismodelling of the beam effect is two-fold.
First, the beam convolution is performed on the sky,
which induces wide-angle effects.
Second, the beam is chromatic, i.e. the beam size is frequency dependent.
Unlike in galaxy surveys where convolutional effects can be expressed
as a window $W(\bm{x} - \bm{y})$ and modelled in the standard formalism,
beam smoothing induces a position-dependent window $W(\bm{x} - \bm{y}| \bm{x})$.
We will further explore the modelling of the beam effect in the future, as mentioned in \secref{sec:future}.

\section{Higher-order multipoles}
\label{apdx:multipoles}
In this section, we introduce and validate the higher-order multipoles of the power spectrum.
The higher-order multipoles are defined as
\begin{equation}
    P_\ell(k) = (2\ell+1)  \langle P(\bm{k},\mu) \mathcal{L}_\ell(\mu) \rangle_{\mu},
\end{equation}
where $\mathcal{L}_\ell(\mu)$ is the Legendre polynomial of order $\ell$ which for the multipoles that we consider are defined as
\begin{equation}\label{eq:legendre polynomials}
    \mathcal{L}_2 = \frac{3\mu^2 - 1}{2}, \hspace{3pt} \mathcal{L}_4 = \frac{35\mu^4 - 30\mu^2 + 3}{8},
\end{equation}
and $\langle \rangle_{\mu}$ denotes the integration over the $\mu$ direction, effectively averaging the $\bm{k}$-modes within the same $k$-bin.
\begin{figure}
    \centering
    \includegraphics[width=\linewidth]{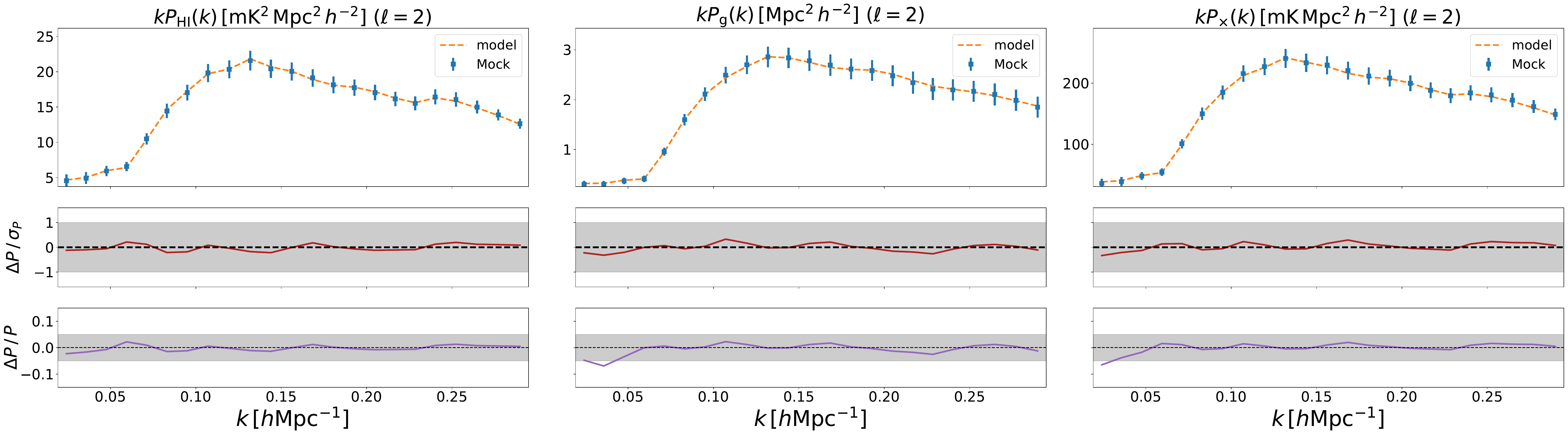}
    \includegraphics[width=\linewidth]{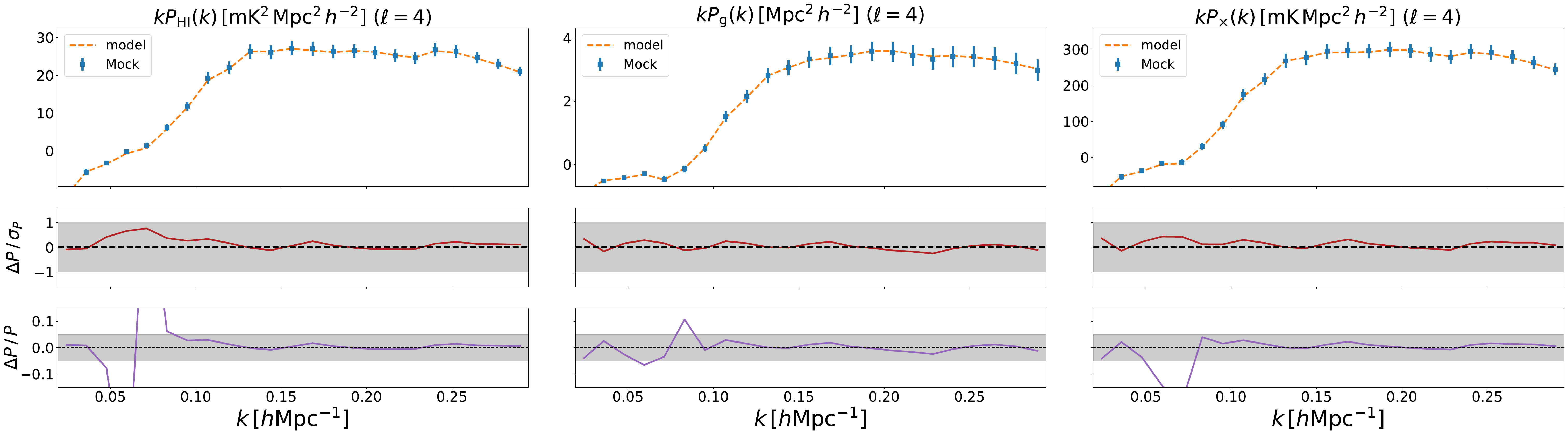}
    \caption{Validation of the higher-order power spectrum multipoles. The top panels in the first row show the 1D quadrupole $(\ell=2)$ and the top panels in the second row show the hexadecapole $(\ell=4)$. The blue squares represent the mock power spectra, and the orange dashed lines represent the input model. The centre and bottom sub-panels are the same as \autoref{fig:estvalidation}, showing the fractional difference between the mock average and the model over the standard deviation with the shaded area denoting the $\pm1\sigma$ region, and the fractional difference between the mock average and the model over the model with the shaded region denoting the $\pm5\%$ region. \github{https://github.com/meerklass/meer21cm/blob/main/papers/validation/multipoles/multipole_validation.ipynb }}
    \label{fig:multipole validation}
\end{figure}
We test the higher-order multipoles following the formalism outlined in \secref{sec:validation}, using the lognormal simulations of the \hi\ temperature field and galaxy positions without including foreground cleaning effects. The results for the quadrupole $(\ell=2)$ and the hexadecapole $(\ell=4)$ are shown in \autoref{fig:multipole validation}. For both multipoles, we test the \hi\ auto-power, the galaxy auto-power, and their cross-power spectrum. The main sub-panels show these 1D power spectrum multipoles, and illustrate the mean mock measurements agreement with the input model power spectra. 

We observe some interesting features in these results. At small $k$, the amplitudes in power are similar to those from the monopole in \autoref{fig:estvalidation}. However, at larger $k$, the amplitude sharply increases, which we mainly attribute to the $|\bm{k}_\perp|\,{>}\,0.06\,{h\,{\rm Mpc}^{-1}}$ scale cut. Excluding large regions of high $\bm{k}_\perp$ effectively restricts the available modes so that $\mu\,{\to}\,1$. This has a more noticeable effect in the higher-order multipoles because they isolate anisotropy by constructing weighted averages over $\mu$ that cancel isotropic contributions via their Legendre polynomial weightings. Although $\mathcal{L}(\mu)\,{\to}\,1$ as $\mu\,{\to}\,1$ for all $\ell$, the impact of restricting to $\mu\,{\sim}\,1$ differs between multipoles. The monopole is a simple average over $\mu$ and does not rely on cancellations, so it is less affected by the loss of modes. In contrast, higher-order multipoles are constructed from weighted averages that involve substantial cancellation between transverse and radial contributions. By removing modes at low $|\mu|$, the cut suppresses these cancellations, leading to an enhancement of the higher-order multipole amplitudes.

Overall, we see excellent agreement between the fiducial model and the mocks, suggesting that our pipeline is suitable for the higher-order multipoles of the power spectra. A crucial next step in continuing the validation of our pipeline on higher-order multipoles will be to test the effects of blind foreground cleaning and examine the accuracy of a foreground transfer function-based reconstruction. Putting these lognormal mocks through a realistic foreground removal scenario will test if higher order multipoles, such as the quadrupole, can be recovered and lay the foundations for the detection of RSD signals with MeerKLASS \citep{detecting_rsd_w_meerklass_inprep}.
In future data analysis, estimators beyond the plane-parallel limits will be explored, as discussed in \secref{sec:future}.

\section{Flagging frequency channels}
\label{apdx:flagging}
\begin{figure}[htbp]
    \centering
    \includegraphics[width=0.5\linewidth]{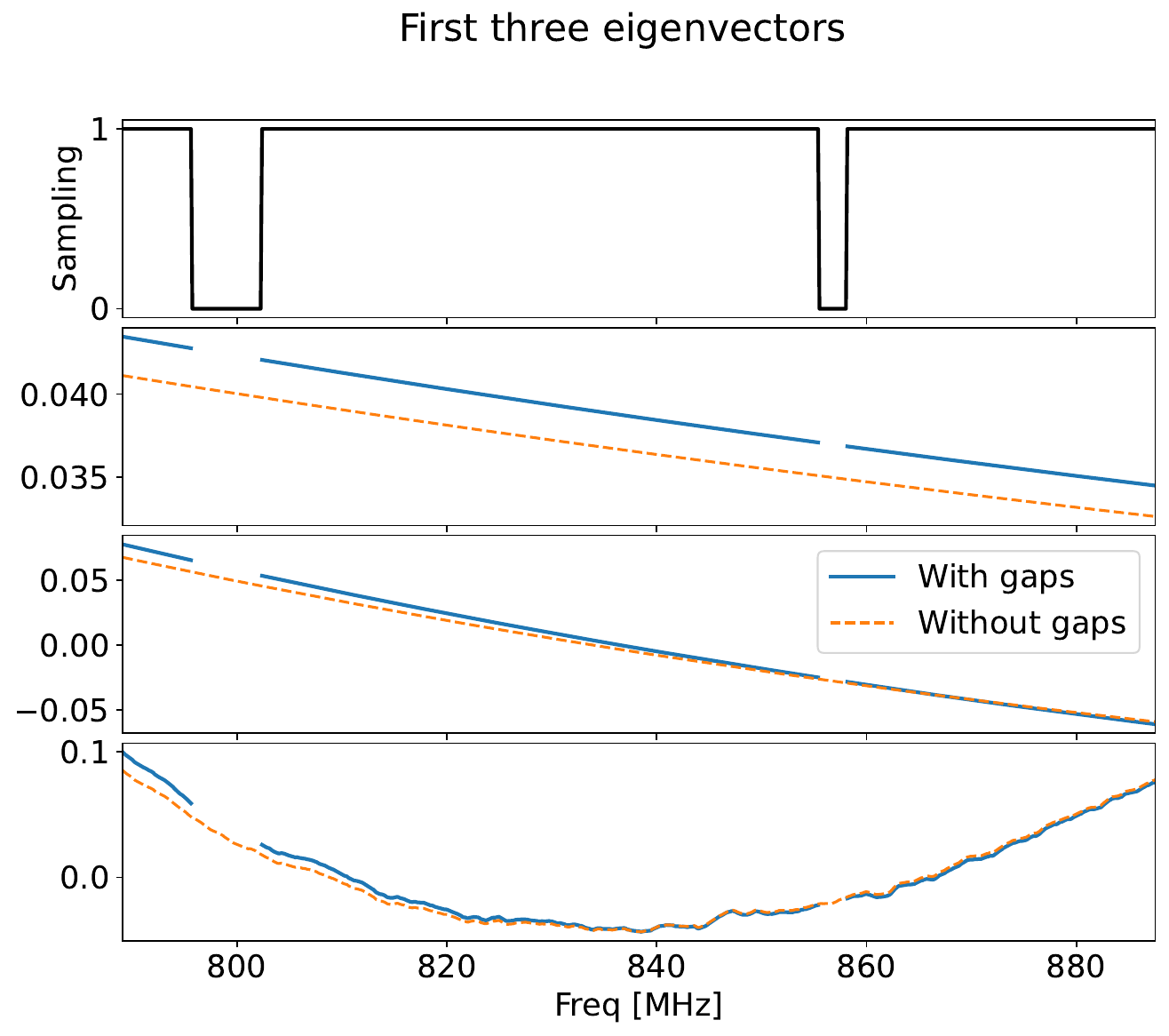}
    \caption{
        The top panel shows the frequency gaps used in \appref{apdx:flagging}.
        The bottom panels show the first three eigenvectors from the PCA cleaning,
        with the blue solid line representing the eigenvectors with frequency gaps,
        and the orange dashed line representing the eigenvectors without frequency gaps.
        \github{https://github.com/meerklass/meer21cm/blob/main/papers/validation/gaps_illus.ipynb}
    }
    \label{fig:flagged_channels}
\end{figure}

\begin{figure}[htbp]
    \centering
    \includegraphics[width=\linewidth]{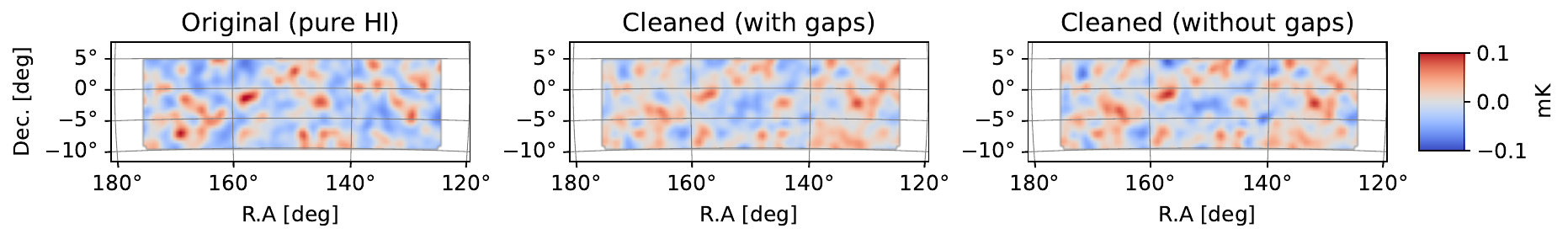}
    \caption{
        The original mock \hi\ map (left panel), 
        the residual map after PCA cleaning with frequency gaps (middle panel), 
        and the residual map after PCA cleaning without frequency gaps (right panel).
        For illustration, the maps shown are the average of the first 20 frequency channels.
        \github{https://github.com/meerklass/meer21cm/blob/main/papers/validation/gaps_illus.ipynb}
    }
    \label{fig:maps_gaps}
\end{figure}

Radio observations are prone to radio frequency interference (RFI) contamination, 
which requires flagging of the observed data, often over entire frequency channels.
As a part of the data calibration process, RFI flagging functionalities are not considered
in the \meer\ pipeline.
Nevertheless, it is important to validate that,
missing frequency channels do not affect the accuracy of the foreground cleaning, power spectrum estimation
and the signal loss correction.

In this section, we give a simple illustration of the robustness of the pipeline to frequency gaps,
by running the mock simulation with completely flagged channels.
In the top panel of \autoref{fig:flagged_channels}, we show the frequency gaps used in the mock simulation.
We mask 20 channels around $\sim 800\,$MHz, and 10 channels around $\sim 855\,$MHz.
We then simulate the same mock \hi\ observation as described in \secref{sec:mock},
but with the frequency gaps applied.
We then perform the PCA cleaning on the masked data, with the eigenvectors shown in the bottom panels of \autoref{fig:flagged_channels}.
We see that the masked data gives the correct recovery of the foreground structure,
with the eigenvectors smoothly varying instead of having step function jumps around the gaps.
The structure is very similar to the case without frequency gaps.
We note that the amplitude of the eigenvectors are slightly different,
simply because the number of channels sampled is different and the eigenvectors are normalised to unit length.

We then compare the residual map after PCA cleaning with and without frequency gaps,
with the results shown in \autoref{fig:maps_gaps}.
We see that the residual map with frequency gaps has almost identical structure
to the case without frequency gaps.
While not shown here, we have verified
that the subsequent power spectrum estimation and transfer function correction are unaffected in accuracy,
as expected from the effective cleaning of the foregrounds.

\section{Analytical modelling of the transfer function}
\label{apdx:pcawindow}

\begin{figure}[b]
    \centering
    \includegraphics[width=\linewidth]{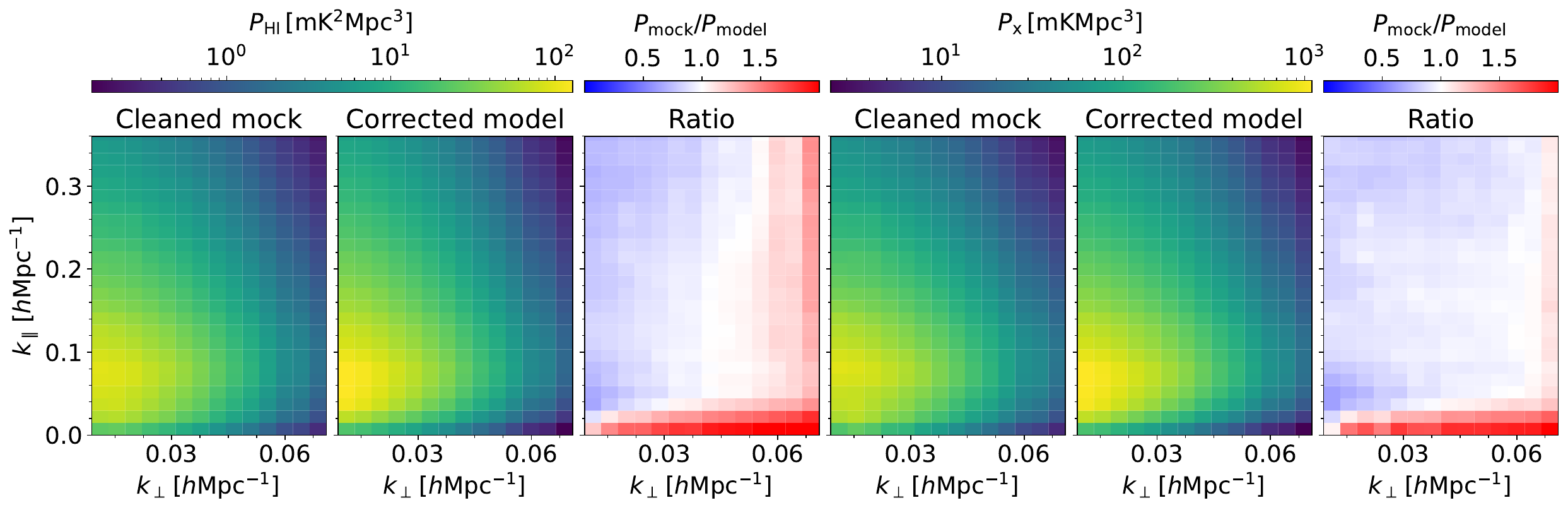}
    \includegraphics[width=\linewidth]{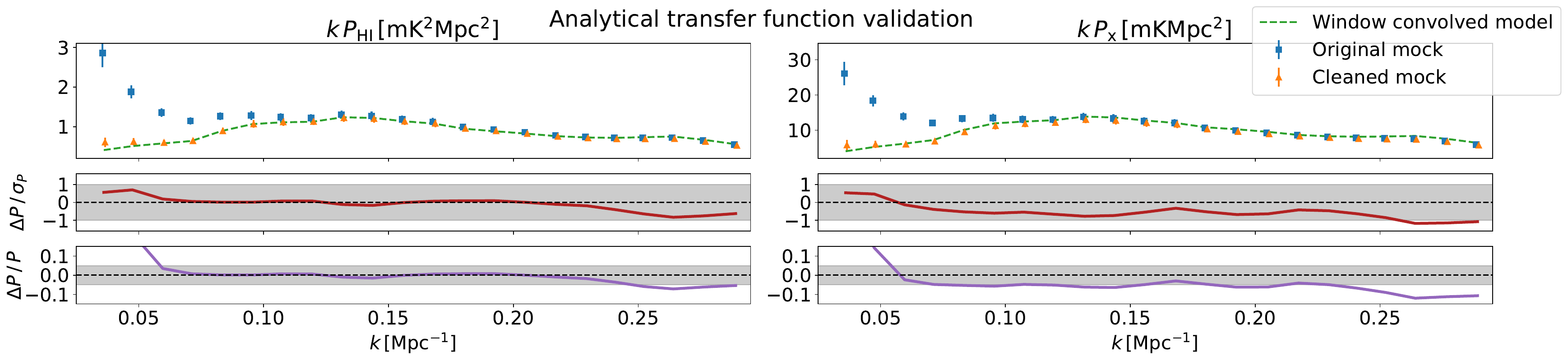}
    \caption{Validation of the analytical modelling of the PCA window function.
    The top left panels show the cylindrical power spectra of the foreground cleaned \hi\ auto-power,
    analytical window function corrected model, and the ratio between the two.
    The top right panels show the same cylindrical power spectra, but for the cross-power.
    The bottom panels show the 1D power spectra, with the blue square representing the original mock power spectrum,
    the orange triangle representing the foreground cleaned mock power spectrum,
    and the green dashed line representing the analytical window function corrected model,
    \github{https://github.com/meerklass/meer21cm/blob/main/papers/validation/plot_02.ipynb}
    }
    \label{fig:pcawindow}
\end{figure}

In this section, we present a preliminary effort to analytically model the PCA effects on the power spectrum,
following the formalism in \cite{2025MNRAS.542L...1C}.
To briefly review, in the flat-sky and plane-parallel approximation, the window function
for the power spectrum estimator arising from the PCA cleaning is given by
\begin{equation}
    \mathbf{H}_{\alpha\beta} = \bigg|(\widetilde{\mathbf{R}}_1)_{\alpha\beta}(\widetilde{\mathbf{R}}_2)^*_{\alpha\beta}\bigg|_{\rm Re},
\end{equation}
where $\widetilde{\mathbf{R}}_1$ and $\widetilde{\mathbf{R}}_2$ are the Fourier transforms of the foreground removal matrix $\mathbf{R}_1$ and $\mathbf{R}_2$, respectively.
The subscript $_{1,2}$ denotes the possibility of cross-correlating different datasets.
In the context of this paper, $\mathbf{R}_1 = \mathbf{R}_2 = \mathbf{R}^{\rm PCA}$ for the case of \hi-auto power.
In the case of \hi-galaxy cross-power, $\mathbf{R}_1 = \mathbf{R}^{\rm PCA}$ and $\mathbf{R}_2 = \mathbf{I}$, where $\mathbf{I}$ is the identity matrix.
As shown in \cite{2025MNRAS.542L...1C}, for cross-power, the window function matrix is a diagonal matrix,
and the correction needed is simply a multiplicative transfer function identical to the model in \autoref{eq:pcrossmodel}.
In the auto-power case, the window function matrix needs to be fully modelled.

If we assume that frequency channels are equally spaced in comoving distance,
and the line-of-sight direction is along the $z$-axis,
then the subscript $\alpha\beta$ denotes different $k_\parallel$ modes.
However, in reality, the frequency channels are not equally spaced in comoving distance,
and the line-of-sight direction is not perfectly along the $z$-axis as illustrated in \autoref{fig:logrealisation_map_grid}.
Therefore, we need to interpolate the window function matrix to be applicable to the actual estimation grid.

To do that, we first assume an effective line-of-sight distance grid for the frequency channels,
\begin{equation}
    \Delta r_\parallel = \big(d_c(\nu_{\rm min}) - d_c(\nu_{\rm max})\big)/N_{\rm ch},
\end{equation}
where $N_{\rm ch}$ is the number of frequency channels.
We can then define the pseudo $k_\parallel$ modes as
\begin{equation}
    k_\parallel^{\rm pseudo} = \frac{2 \pi\,n}{N_{\rm ch} \Delta r_\parallel}, \quad n = 0, 1, 2, \ldots, N_{\rm ch}-1,
\end{equation}
so that the indices of the window function matrix $\mathbf{H}_{\alpha\beta}$ is simply $(k_\parallel^{\rm pseudo})_{\alpha},(k_\parallel^{\rm pseudo})_{\beta}$.
We can then calculate the signal loss as a function of the pseudo $k_\parallel$ modes,
\begin{equation}
    \mathcal{T}\big((k_\parallel^{\rm pseudo})_{\alpha}\big) = \sum_\beta \mathbf{H}_{\alpha\beta},
\end{equation}
which we then interpolate to the actual $k_\parallel$ modes.
This interpolated transfer function is then multiplied to the model cross-power spectrum to correct for the signal loss,
\begin{equation}
    P_{\times}^{\rm corrected}(\bm{k}) = P_{\times}^{\rm model}(\bm{k}) \times \mathcal{T}\big(k_\parallel).
\end{equation}

For the auto-power, we instead perform an interpolation on $\mathbf{H}_{\alpha\beta}$ itself
to the actual $(k_\parallel, k_\parallel)$ grids.
Note that, the interpolation does not preserve the normalisation.
For the same mode $(k_\parallel^{\rm pseudo})_{\alpha} = (k_\parallel)_{i}$,
the interpolated window function $\mathbf{H}_{ij}^{\rm interp}$ needs to be normalised to match the original window function,
\begin{equation}
    \mathbf{H}_{ij} = \frac{\mathbf{H}_{ij}^{\rm interp}}{\sum_k \mathbf{H}_{ik}^{\rm interp}} \mathcal{T}\big((k_\parallel)_{i}\big).
\end{equation}
The final window function matrix is then convolved with the model power spectrum along the $k_\parallel$ direction.

The results are shown in \autoref{fig:pcawindow}.
We see that the analytical window function corrected model 
roughly matches the cleaned mock power spectrum, for both the auto- and cross-power.
However, the accuracy is not perfect, with the corrected model
overpredicting the power at higher $k_\parallel$ and underpredicting at lower $k_\parallel$.
This is likely due to the noisy interpolation of the window function,
coupled with the fact that the PCA cleaning is not exactly along the $z$ direction,
unlike the global plane-parallel approximation in the modelling.
In the large-scale limit, where the wide-angle and the PCA signal loss effects are prominent,
the analytical model is particularly prone to the inaccuracies.
Further studies into better modelling of the window function beyond
current approximations will be conducted in future works.

Nevertheless, the analytical correction still provides a good recovery at
the level of 1D monopole power spectrum as we show in the bottom panels of \autoref{fig:pcawindow}.
As one can see, the accuracy between $k\sim 0.05\,{h{\rm Mpc}^{-1}}$ to $k\sim 0.1\,{h{\rm Mpc}^{-1}}$,
where signal loss is significant, is within $1\,\sigma$ level and reaches per-cent level precision.
This is likely due to some cancellation of the over- and under-predictions when performing the 1D average.
The accuracy at larger scales $k<0.05\,{h{\rm Mpc}^{-1}}$ is poorer,
which is expected given the discussions on the cylindrical power spectra.

Overall, the analytical modelling of the window function provides a
promising alternative to the numerical transfer function correction.
If more accurate modelling of the window function can be achieved,
future data analysis will be able to use it for parameterising and
marginalising over the PCA cleaning effects, as well as building
analytical covariance estimation.
The numerical transfer function can also be extended to model the off-diagonal power spectrum between different $\bm{k}$-modes to reconstruct the window function matrix, using the framework discussed in this Appendix.

\section{Parameter inference}
\label{apdx:parameterinference}

\begin{table}[htbp]
    \centering
    \begin{tabular}{c|c|c|c|c|c|c}
       Notation  & $\log_{10}[\Omega_{\rm \hi}]$ &  $b_{\rm \hi}$ & $b_{\rm gal}$ & $r_\times$ & $\sigma_{\rm p}^{\rm \hi}$ & $\sigma_{\rm p}^{\rm gal}$ \\
       \hline
       Prior  & $[-6, 0]$ & [0.1, 10.0] & [0.1, 10.0] & [0.0, 1.0] & $[0, 2000]\,[{\rm km/s}]$ & $[0, 2000]\,[{\rm km/s}]$ \\
       \hline
       \noalign{\vskip 0.5em}
       Posterior & ${-3.24}_{-0.29}^{+0.18}$ & ${0.77}_{-0.49}^{+1.31}$ & ${1.03}_{-0.11}^{+0.11}$ & ${0.77}_{-0.21}^{+0.12}$ & ${138.94}_{-92.23}^{+109.25}$ & ${154.81}_{-104.07}^{+129.14}$ \\
    \end{tabular}
    \caption{Prior and posterior distributions used for the parameter inference. 
    The first row shows the lower and upper limits of the uniform prior distributions.
    The second row shows the posterior distributions, including the median value and $68\%$ confidence interval.}
    \label{tab:parameters}
\end{table}

\begin{figure}[htbp]
    \centering
    \includegraphics[width=\linewidth]{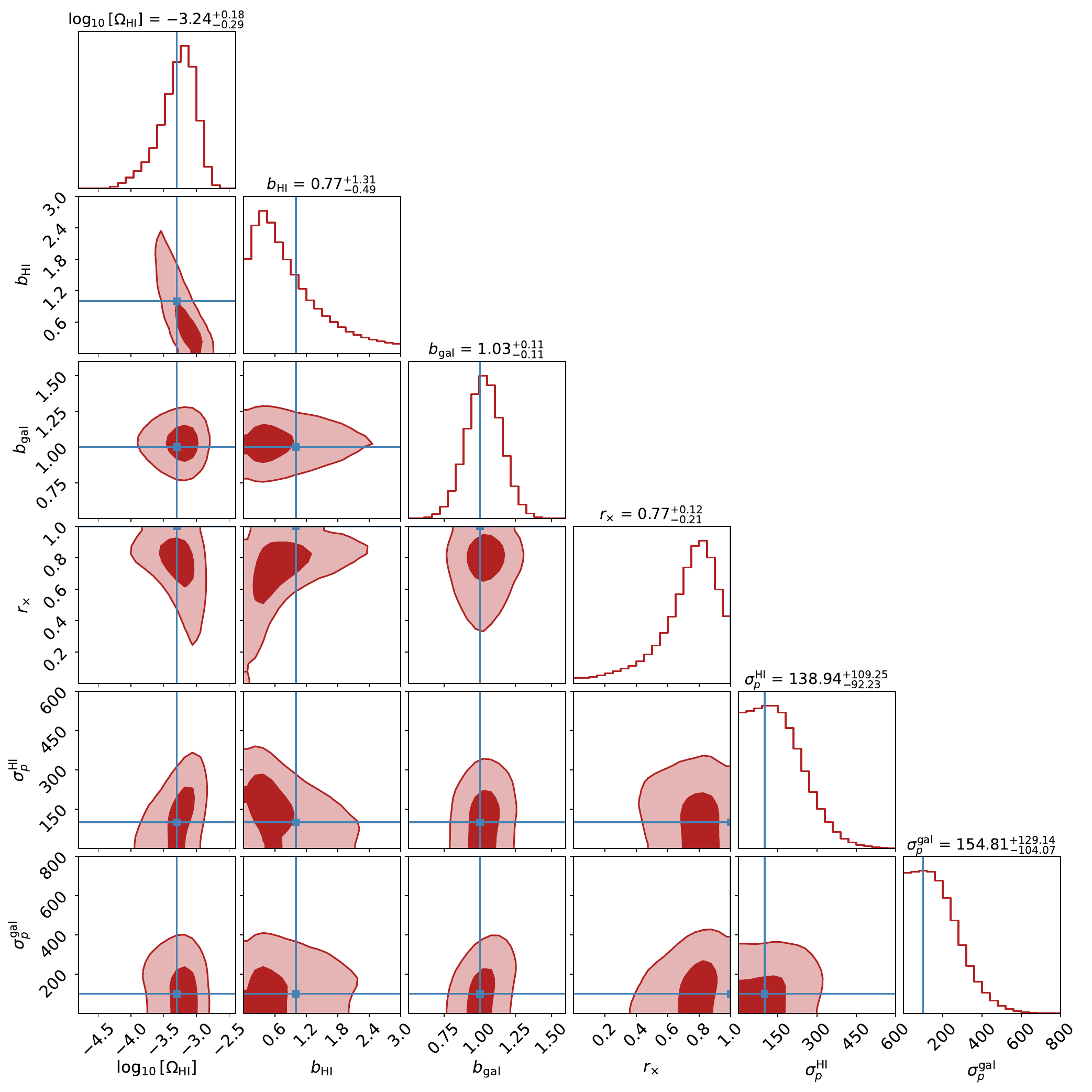}
    \caption{
        Posterior distribution of the tracer-dependent parameters inferred from the mock data and covariance.
        The inner/outer contours show the $1\sigma$/$2\sigma$ confidence regions.
        The fiducial parameter values are marked with the blue squares.
        The title of each histogram shows the median and $68\%$ confidence interval of the posterior.
        \github{https://github.com/meerklass/meer21cm/blob/main/papers/validation/fitting.ipynb}
    }
    \label{fig:posterior}
\end{figure}

We have demonstrated that unbiased estimation can be performed with \meer, as shown in \secref{sec:validation}.
Given the monopole power spectrum and its covariance generated through the mock realisations,
we can then perform parameter inference using the \meer\ pipeline following \secref{subsec:modelinference}.

For illustration, we assume fixed cosmology and vary the tracer-dependent parameters,
including $[\Omega_{\rm \hi}, b_{\rm \hi}, b_{\rm gal}, r_\times, \sigma_{\rm p}^{\rm \hi}, \sigma_{\rm p}^{\rm gal}]$.
We fit the multi-tracer data vector containing the \hi\ auto-power, galaxy auto-power and the cross-power, with the covariance
illustrated in \autoref{fig:corr} in \secref{sec:validation}.
The log-likelihood function is given by \autoref{eq:loglike}.
We use wide, uniform prior distributions for the parameters, as shown in the first row of \autoref{tab:parameters}.
The sampling is performed using the \textsc{nautilus} package \citep{2023MNRAS.525.3181L},
and we provide an intuitive interface in \meer\ for running the sampler.

The results are shown in \autoref{fig:posterior}.
We see that the posterior distributions are consistent with the fiducial parameters,
except for the cross-correlation coefficient $r_\times$, which is slightly biased towards lower values.
This is because the cross-correlation power spectrum is slightly over-estimated by the model, as seen in \autoref{fig:tfvalidation}.
Furthermore, since $r_\times$ is physically bounded between 0 and 1, the posterior distribution is truncated near the fiducial value of 1.
We verify that allowing unphysical values of $r_\times$ will result in a $1\sigma$ confidence region that includes the fiducial value.
The mean of the posterior for the \hi\ bias parameter, $b_{\rm \hi}$, is slightly biased while the deviation is well within the $1\sigma$ confidence interval.

Our results demonstrate that, when assuming fixed cosmology, the degeneracy between the bias parameters
and the mean \hi\ brightness temperature can be broken, resulting in unbiased estimation of the properties.
In the future, measurements of the power spectra from MeerKLASS UHF survey data together with DESI
will shed light on the physical properties of the \hi\ and the galaxy-halo connection, through constraining these parameters.
For future surveys with high signal-to-noise,
jointly modelling the cosmology and the tracer-dependent parameters will be explored,
as discussed in \secref{sec:future}.

\bibliographystyle{mnras}
\bibliography{Bib} 

\end{document}